\documentclass[12pt]{article}

\usepackage{xr-hyper}
\usepackage{amsmath, amsfonts, amssymb, amsthm, bbm, enumerate, titlesec, graphicx, epstopdf, multirow, array, xcolor, bbm, adjustbox}
\usepackage[T1]{fontenc}
\usepackage{lmodern}
\usepackage[title]{appendix}

\usepackage{fullpage}
\usepackage[onehalfspacing]{setspace}

\titleformat*{\paragraph}{\sc}

\definecolor{darkred}{rgb}{0.5,0,0}
\usepackage{hyperref}
\hypersetup{allbordercolors={1 1 1},allcolors=darkred,colorlinks=true}
\usepackage[capitalize,nameinlink,noabbrev]{cleveref}

\usepackage[small,bf]{caption}

\usepackage[round]{natbib}
\usepackage[runin]{abstract}
\newcolumntype{C}[1]{>{\centering\arraybackslash}p{#1}}

\theoremstyle{plain}
\newtheorem{asn}{Assumption}
\crefname{asn}{Assumption}{Assumptions}

\crefname{lem}{Lemma}{Lemmas}
\newtheorem{thm}{Theorem}

\newtheorem{cor}{Corollary}

\newtheorem*{claim*}{Claim}

\allowdisplaybreaks

\usepackage{tikz}
\usetikzlibrary{arrows,positioning}
\tikzset{
    >=stealth',
    punkt/.style={
           rectangle,
           rounded corners,
           draw=black, very thick,
           text centered,
           text width=1.5cm,
           minimum height=20pt,
           text depth=0pt},
    pil/.style={
           ->,
           thick,
           shorten <=4pt,
           shorten >=4pt}
}

\crefname{sappsec}{Supplemental Appendix}{Supplemental Appendices}
\crefname{sappsubsec}{Supplemental Appendix}{Supplemental Appendices}
\crefname{sappsubsubsec}{Supplemental Appendix}{Supplemental Appendices}
\crefname{appsec}{Appendix}{Appendices}
\begin{document}

\title{Full-Information Estimation of Heterogeneous Agent Models Using Macro and Micro Data}

\author{
        \begin{tabular}{ccc}
        Laura Liu & & Mikkel Plagborg-M{\o}ller\thanks{{\tt lauraliu@iu.edu} (Liu) and {\tt mikkelpm@princeton.edu} (Plagborg-M{\o}ller). We are grateful for helpful comments from two anonymous referees, Adrien Auclert, Yoosoon Chang, Marco Del Negro, Simon Mongey, Hyungsik Roger Moon, Ulrich M\"{u}ller, Jonathan Payne, Frank Schorfheide, Neil Shephard, Thomas Winberry, Tao Zha, and participants at various seminars and conferences. Plagborg-M{\o}ller acknowledges that this material is based upon work supported by the NSF under Grant {\#}1851665. Any opinions, findings, and conclusions or recommendations expressed in this material are those of the authors and do not necessarily reflect the views of the NSF.} \\[-4pt]
        {\em Indiana University} && {\em Princeton University}
        \date{First version: December 19, 2018 \texorpdfstring{\\}{} This version: \today}
        \end{tabular}
}

\date{\today}
\maketitle


\begin{abstract}
We develop a generally applicable full-information inference method for heterogeneous agent models, combining aggregate time series data and repeated cross sections of micro data. To handle unobserved aggregate state variables that affect cross-sectional distributions, we compute a numerically unbiased estimate of the model-implied likelihood function. Employing the likelihood estimate in a Markov Chain Monte Carlo algorithm, we obtain fully efficient and valid Bayesian inference. Evaluation of the micro part of the likelihood lends itself naturally to parallel computing.
Numerical illustrations in models with heterogeneous households or firms demonstrate that the proposed full-information method substantially sharpens inference relative to using only macro data, and for some parameters micro data is essential for identification.
\end{abstract}


\section{Introduction}
Macroeconomic models with heterogeneous agents have exploded in popularity in recent years.\footnote{For references and discussion, see \citet{Krueger2016}, \citet{Ahn2017}, and \citet{Kaplan2018JEP}.} New micro data sets -- including firm and household surveys, social security and tax records, and censuses -- have exposed the empirical failures of traditional representative agent approaches. The new models not only improve the fit to the data, but also make it possible to meaningfully investigate the causes and consequences of inequality among households or firms along several dimensions, including endowments, financial constraints, age, size, location, etc.

So far, however, empirical work in this area has only been able to exploit limited features of the micro data sources that motivated the development of the new models. As emphasized by \citet{Ahn2017}, the burgeoning academic literature has mostly calibrated model parameters and performed over-identification tests by matching a few empirical moments that are deemed important \emph{a priori}. This approach may be highly inefficient, as it ignores that the models' implied macro dynamics and cross-sectional properties often fully determine the entire \emph{distribution} of the observed macro and micro data. The failure to exploit the joint information content of macro and micro data stands in stark contrast to the well-developed inference procedures for estimating \emph{representative} agent models using only macro data \citep{Herbst2016}. 

To exploit the full information content of macro and micro data, we develop a general technique to perform Bayesian inference in heterogeneous agent models. We assume the availability of aggregate time series data as well as repeated cross sections of micro data. Evaluation of the joint macro and micro likelihood function is complicated by the fact that the model-implied cross-sectional distributions typically depend on unobserved aggregate state variables. To overcome this problem, we devise a way to compute a numerically unbiased estimate of the model-implied likelihood function of the macro and micro data. As argued by \citet{Andrieu2010} and \citet{Flury2011}, such an unbiased likelihood estimate can be employed in standard Markov Chain Monte Carlo (MCMC) procedures to generate draws from the fully efficient Bayesian posterior distribution given all available data.

The starting point of our analysis is the insight that existing solution methods for heterogeneous agent models directly imply the functional form of the joint sampling distribution of macro and micro data, given structural parameters. These models are typically solved numerically by imposing a flexible functional form on the relevant cross-sectional distributions (e.g., a discrete histogram or parametric family of densities). The distributions are governed by time-varying unobserved state variables (e.g., moments). To calculate the model-implied likelihood, we decompose it into two parts. First, heterogeneous agent models are typically solved using the method of \citet{Reiter2009}, which linearizes with respect to the macro shocks but not the micro shocks. Hence, the \emph{macro part} of the likelihood can be evaluated using standard linear state space methods, as proposed by \citet{Mongey2017} and \citet{Winberry2018}.\footnote{If non-Reiter model solution methods are used, our general estimation approach could in principle still be applied, though its computational feasibility would be context-dependent, as discussed in \cref{sec:concl}.} Second, the likelihood of the repeated cross sections of \emph{micro data, conditional on the macro state variables}, can be evaluated by simply plugging into the assumed cross-sectional density. The key challenge that our method overcomes is that the econometrician typically does not directly observe the macro state variables. Instead, the observed macro time series are imperfectly informative about the underlying states.

Our procedure can loosely be viewed as a rigorous Bayesian version of a two-step approach: First we estimate the latent macro states from macro data, and then we compute the model-implied cross-sectional likelihood conditional on these estimated macro states. More precisely, we obtain a \emph{numerically unbiased} estimate of the likelihood by averaging the cross-sectional likelihood across repeated draws from the smoothing distribution of the hidden states given the macro data. We emphasize that, despite being based on a likelihood estimate, our method is fully Bayesian and automatically takes into account all sources of uncertainty about parameters and states. An attractive computational feature is that evaluation of the micro part of the likelihood lends itself naturally to parallel computing. Hence, computation time scales well with the size of the data set. Though our baseline method is designed for repeated cross sections of micro data, we present ideas for exploiting panel data in \cref{sec:panel}.

We perform finite-sample valid and fully efficient Bayesian inference by plugging the unbiased likelihood estimate into a standard MCMC algorithm. The generic arguments of \citet{Andrieu2010} and \citet{Flury2011} imply that the ergodic distribution of the MCMC chain is the full-information posterior distribution that we would have obtained if we had known how to evaluate the exact likelihood function (not just an unbiased estimate of it). This is true no matter how many smoothing draws are used to compute the unbiased likelihood estimate. In principle, we may use any MCMC posterior sampling algorithm that relies only on evaluating (the unbiased estimate of) the posterior density, such as Random Walk Metropolis-Hastings.

In contrast to other estimation methods, our full-information method is automatically finite-sample efficient and can easily handle unobserved individual heterogeneity, micro measurement error, as well as data imperfections such as selection or censoring. In an important early work, \citet{Mongey2017} propose to exploit micro data by collapsing it to time series of cross-sectional moments and incorporating these into the macro likelihood. \emph{In principle}, this approach can be as efficient as our full-information approach if the structural model implies that these moments are sufficient statistics for the micro data. We provide examples where this is not the case, for example due to the presence of unobserved individual heterogeneity and/or micro measurement error. Even when sufficient statistics do exist, it is necessary to properly account for sampling error in the observed cross-sectional moments, which is done automatically by our full-information likelihood method, but could be delicate and imprecise for moment-based approaches. Moreover, textbook adjustments to the micro likelihood allow us to accommodate specific empirically realistic features of micro data such as selection (e.g., over-sampling of large firms) or censoring (e.g., top-coding of income), whereas this is challenging to do efficiently with moment-based approaches.

We illustrate the joint inferential power of macro and micro data through two numerical examples: a heterogeneous household model \citep{Krusell1998} and a heterogeneous firm model \citep{Khan2008}. In both cases we assume that the econometrician observes certain standard macro time series as well as intermittent repeated cross sections of, respectively, (i) household employment and income and (ii) firm capital and labor inputs. Using simulated data, and given flat priors, we show that our full-information method accurately recovers the true structural model parameters. Importantly, for several structural parameters, the micro data reduces the length of posterior credible intervals substantially, relative to inference that exploits only the macro data. In fact, we give examples of parameters that can only be identified if micro data is available. In contrast, inference from moment-based approaches can be highly inaccurate and sensitive to the choice of moments.

We deliberately keep our numerical illustrations low-dimensional and build our code on top of the user-friendly Dynare-based model solution method of \citet{Winberry2018}. Though pedagogically useful, this particular numerical model solution method cannot handle very rich models, so a full-scale empirical illustration is outside the scope of this paper. However, there is nothing in our general inference approach that rules out larger-scale models. We argue in \cref{sec:concl} that our general inference approach is compatible with cutting-edge model solution methods that apply automatic dimension reduction of the state space equations \citep{Ahn2017}.

\paragraph{Literature.}
Our paper contributes to the recent literature on structural estimation of heterogeneous agent models by exploiting the full, combined information content available in macro and micro data. We build on the idea of \citet{Mongey2017} and \citet{Winberry2018} to estimate heterogeneous agent models from the linear state space representation obtained from the \citet{Reiter2009} model solution approach. Several papers have exploited only macro data (as well as calibrated steady-state micro moments) for estimation, including \citet{Winberry2018}, \citet{Hasumi2019}, \citet{Acharya2020}, \citet{Auclert2020humps}, and \citet{Auclert2020seq}. \citet{Challe2017}, \citet{Mongey2017}, \citet{Bayer2020}, and \citet{Papp2020} additionally track particular cross-sectional moments over time. In contrast, we exploit the entire model-implied likelihood function given repeated micro cross sections, which is (at least weakly) more efficient, as discussed further in \cref{sec:theory_moment}.

We are not aware of other papers that tackle the fundamental problem that the aggregate shocks affecting cross-sectional heterogeneity are not directly observed. \citet{ParraAlvarez2020} use the model-implied steady-state micro likelihood in a heterogeneous household model, but abstract from macro data or aggregate dynamics. Closest to our approach are \citet{FernandezVillaverde2018}, who exploit the model-implied joint sampling density of macro and micro data in a particular heterogeneous agent macro model. However, they assume that the underlying state variables are directly observed, whereas our contribution is to solve the computational challenges that arise in the generic case where the macro states are (partially) latent.

Certain other existing methods for combining macro and micro data cannot be applied in our setting. \citet{Hahn2018} develop asymptotic theory for estimation using interdependent micro and macro data sets, but their full-information approach requires derivatives of the exact likelihood in closed form, which is not available in our setting due to the need to integrate out unobserved state variables. \citet{Chang2018} propose a \emph{reduced-form} approach to estimating the feedback loop between aggregate time series and heterogeneous micro data; they do not consider estimation of structural models. In likelihood estimation of \emph{representative} agent models, micro data has mainly been used to inform the prior, as in \citet{Chang2002}. Finally, unlike the microeconometric literature on heterogeneous agent models \citep{Arellano2017}, our work explicitly seeks to estimate the deep parameters of a general equilibrium macro model by also incorporating aggregate time series data.

\paragraph{Outline.}
\cref{sec:model} shows that heterogeneous agent models imply a fully-specified statistical model for the macro and micro data. \cref{sec:method} presents our method for computing an unbiased likelihood estimate that is used to perform efficient Bayesian inference. There we also compare our full-information approach with moment-based estimation approaches. \cref{sec:example_hh,sec:example_firm} illustrate the inferential power of combining macro and micro data using two simple numerical examples, a heterogeneous household model and a heterogeneous firm model. \cref{sec:panel} proposes an extension to panel data. \cref{sec:concl} concludes and discusses possible future research directions. \cref{sec:appendix} contains proofs and technical results. A Supplemental Appendix and a full Matlab code suite are available online.\footnote{\url{https://github.com/mikkelpm/het_agents_bayes}}

\section{Framework}
\label{sec:model}
We first describe how heterogeneous agent models generically imply a statistical model for the macro and micro data. Then we illustrate how a simple model with heterogeneous households fits into this framework.

\subsection{A general heterogeneous agent framework}
\label{sec:framework}
\begin{figure}[tp]
\centering
\begin{tikzpicture}[node distance=1cm, auto]

 \node[punkt] (z_tm1) {$z_{t-1}$};
 \node[punkt, right=3cm of z_tm1] (z_t) {$z_t$};
 \node[punkt, right=3cm of z_t] (z_tp1) {$z_{t+1}$};
 \node[left=0.3cm of z_tm1] (ldots) {\dots};
 \node[right=0.3cm of z_tp1] (rdots) {\dots};
 
 \draw[pil] (z_tm1) -- node [midway,above=1mm] {\small $p(z_t \mid z_{t-1},\theta)$} (z_t);
 \draw[pil] (z_t) -- (z_tp1);
 
 \node[below left=1.2cm and 0.6cm of ldots] (lline) {};
 \node[below right=1.2cm and 0.6cm of rdots] (rline) {};
 \draw[dashed] (lline) -- (rline);

 \node[above right=-0.1cm of lline,font=\bf] {$\uparrow$ latent}; 
 \node[below right=-0.1cm of lline,font=\bf] {$\downarrow$ observed};
 
 \node[punkt, below right=2.8cm and 0.1cm of z_tm1] (x_t) {$x_t$};
 \node[punkt, below left=2.8cm and 0.1cm of z_tp1, text depth=2pt] (y_it) {$\lbrace y_{i,t} \rbrace_{i=1}^{N_t}$};
 
 \draw[pil] (z_t) -- node [pos=0.7,left=3mm] {\small $p(x_t \mid z_t,\theta)$} (x_t);
 \draw[pil] (z_t) -- node [pos=0.7,right=4mm] {\small $p(y_{i,t} \mid z_t,\theta)$ \footnotesize (i.i.d.\ across $i$)} (y_it);
 
 \node[right=0.1cm of y_it] {\footnotesize (observed for $t \in \mathcal{T}$)};
 
\end{tikzpicture}
\caption{Diagram of the distribution of the macro and micro data implied by a heterogeneous agent model. The state vector $z_t$ includes any time-varying parameters that govern the cross-sectional distribution $p(y_{i,t} \mid z_t,\theta)$.}
\label{fig:model}
\end{figure}
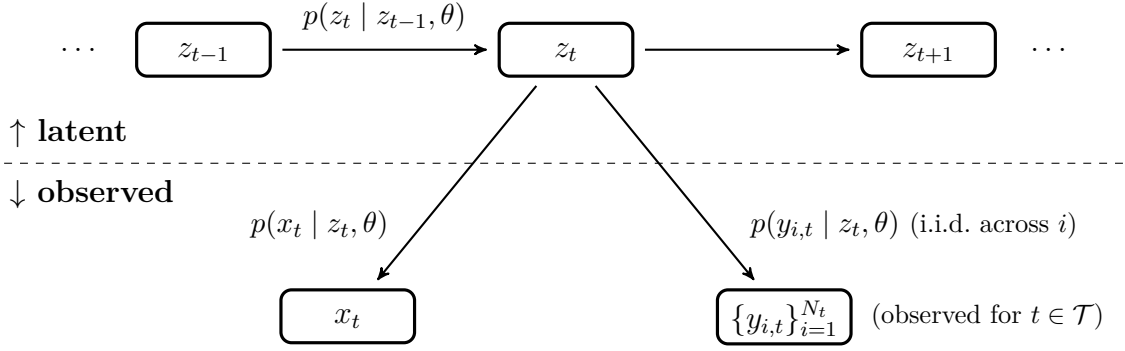

Consider a given structural model that implies a fully-specified equilibrium relationship among a set of aggregate and idiosyncratic variables. We assume the availability of macro time series data as well as repeated cross sections of micro data, as summarized in \cref{fig:model}. Let $\mathbf{x} \equiv \lbrace x_t \rbrace_{1 \leq t \leq T}$ denote the vector of observed time series data (e.g., real GDP growth), where $x_t$ is a vector, and $T$ denotes the time series sample size. At a subset $\mathcal{T} \subset \lbrace 1,2,\dots,T \rbrace$ of time points we additionally observe the micro data $\mathbf{y} \equiv \lbrace y_{i,t} \rbrace_{1 \leq i \leq N_t, t \in \mathcal{T}}$, where $y_{i,t}$ is a vector (e.g., the asset holdings of household $i$ or the employment of firm $i$). At each time $t$, the cross section $\lbrace y_{i,t} \rbrace_{1 \leq i \leq N_t}$ is sampled at random from the model-implied cross-sectional distribution \emph{conditional} on some macro state vector $z_t$. For now it is convenient to assume that $\lbrace y_{i,t} \rbrace$ constitutes a representative sample, but sample selection or censoring are easily accommodated in the framework, as we demonstrate in \cref{sec:example_firm_censor}. Formally, we make the following assumption.
\begin{asn} \label{asn:model}
The data is sampled as follows:
\begin{enumerate}
\item Conditional on $\mathbf{z} \equiv \lbrace z_t \rbrace_{t=1}^T$, the micro data $\lbrace y_{i,t} \rbrace_{1 \leq i \leq N_t, t \in \mathcal{T}}$ is independent across $t$ and the data points $\lbrace y_{i,t} \rbrace_{i=1}^{N_t}$ at time $t$ are sampled i.i.d.\ from the density $p(y_{i,t} \mid z_t,\theta)$.
\item Conditional on $\mathbf{z}$, the micro data $\mathbf{y}$ is independent of the macro data $\mathbf{x}$.
\item Conditional on $z_t$ and $\lbrace x_\tau,z_\tau \rbrace_{\tau \leq t-1}$, the macro data $x_t$ is sampled from the density $p(x_t \mid z_t,\theta)$. Conditional on $\lbrace z_\tau \rbrace_{\tau \leq t-1}$, the state vector $z_t$ is sampled from the density $p(z_t \mid z_{t-1},\theta)$.
\end{enumerate}
\end{asn}
The first condition above operationalizes the notion of representative sampling of repeated cross sections. The second condition entails no loss of generality, since we can always include $x_t$ in the state vector $z_t$. The third condition is a standard Markovian state space formulation of the aggregate dynamics, as discussed further below.

Given the structural parameter vector $\theta$, the fully-specified heterogeneous agent model implies functional forms for the macro observation density $p(x_t \mid z_t,\theta)$, the macro state transition density $p(z_t \mid z_{t-1},\theta)$, and the micro sampling density $p(y_{i,t} \mid z_t,\theta)$. These density functions reflect the equilibrium of the model, as we illustrate in the next subsection, and they are the key inputs in the likelihood computation in \cref{sec:method}. Notice that the framework allows the micro and macro data to be dependent, though this dependence must be fully captured by the macro state vector $z_t$, which is determined by the structure of the model at hand. Because the sampling densities $p(x_t \mid z_t,\theta)$ and $p(y_{i,t} \mid z_t,\theta)$ are derived from an equilibrium model, the likelihood function derived below in equation \eqref{eqn:micro_likel_int} automatically embodies any constraints of the type envisioned by \citet{Imbens1994} on the relationship between the aggregate macro data and the time-varying population moments of the micro sampling distribution. For example, if $y_{i,t}$ equals individual-level consumption, $x_t$ equals aggregate consumption, and $z_t$ equals the underlying macro shocks (which determine the dynamics of aggregates and of the micro distribution), then the asymptotic adding-up constraint that $\lim_{N_t \to \infty}\frac{1}{N_t} \sum_{i=1}^{N_t} y_{i,t} \stackrel{a.s.}{=} x_t$ will be automatically satisfied if the sampling densities are derived from a model that imposes market clearing.

In most applications, some of the aggregate state variables $z_t$ that influence the macro and micro sampling densities are unobserved, i.e., $z_t \neq x_t$. This fact complicates the evaluation of the exact likelihood function and is the key technical challenge that we overcome in this paper, as discussed in \cref{sec:method}.

\subsection{Example: Heterogeneous household model}
\label{sec:model_hh}
We use a simple heterogeneous household model \`{a} la \citet{Krusell1998} to illustrate the components of the general framework introduced in \cref{sec:framework}. Our discussion of the model and the numerical equilibrium solution technique largely follows \citet{Winberry2016,Winberry2018}. Though this model is far too stylized for quantitative empirical work, we demonstrate the flexibility of our framework by adding complications such as permanent heterogeneity among households as well as measurement error in observables. In \cref{sec:example_hh} we will estimate a calibrated version of this model on simulated data.

\paragraph{Model assumptions.}
A continuum of heterogeneous households $i \in [0,1]$ are exposed to idiosyncratic employment risk as well as aggregate shocks to wages and asset returns. Households have log preferences over consumption $c_{i,t}$ at time $t=0,1,2,\dots$. When employed ($\epsilon_{i,t}=1$), households receive wage income net of an income tax levied at rate $\tau$. When unemployed ($\epsilon_{i,t}=0$), they receive unemployment benefits equal to a fraction $b$ of their hypothetical working wage. The idiosyncratic unemployment state $\epsilon_{i,t}$ evolves exogenously according to a two-state first-order Markov process that is independent of aggregate conditions and household decisions. Households cannot insure themselves against their employment risk, since the only available financial instruments are shares of capital $\tilde{a}_{i,t}$, which yield a rate of return $r_t$. Financial investment is subject to the borrowing constraint $\tilde{a}_{i,t} \geq 0$.

For expositional purposes, we add a dimension of permanent household heterogeneity: Each household is endowed with a permanent labor productivity level $\lambda_i$, which is drawn at the beginning of time from a lognormal distribution with mean parameter $\mathbb{E}[\log \lambda_i] = \mu_\lambda \leq 0$ and variance parameter chosen such that $\mathbb{E}[\lambda_i]=1$. An employed household inelastically supplies $\lambda_i$ efficiency units of labor, earning pre-tax income of $\lambda_i w_t$, where $w_t$ is the real wage per efficiency unit of labor.

To summarize, the households' problem can be written
\begin{align*}
&\max_{c_{i,t},a_{i,t} \geq 0} \mathbb{E}_0\left[\sum_{t=0}^\infty \beta^t \log c_{i,t} \right] \\
&\;\text{s.t.} \quad c_{i,t} = \lambda_i\big\lbrace w_t[(1-\tau)\epsilon_{i,t} + b(1-\epsilon_{i,t})] + (1+r_t)a_{i,t-1} - a_{i,t} \big\rbrace,
\end{align*}
where $a_{i,t}=\tilde{a}_{i,t}/\lambda_i$ are the normalized asset holdings.

A representative firm produces the consumption good using a Cobb-Douglas production function $Y_t=e^{\zeta_t}K_t^\alpha L^{1-\alpha}$, where aggregate capital $K_t$ depreciates at rate $\delta$, and $L$ is the aggregate level of labor efficiency units (which is constant over time since employment risk is purely idiosyncratic). The firm hires labor and rents capital in competitive input markets. Log total factor productivity (TFP) evolves as an AR(1) process $\zeta_t = \rho_\zeta \zeta_{t-1} + \varepsilon_t$, where $\varepsilon_t \overset{i.i.d.}{\sim} N(0,\sigma_\zeta^2)$. The government balances its budget period by period, implying $\tau L = b(1-L)$.

We collect the deep parameters of this model in the vector $\theta$. These include $\beta$, $\alpha$, $\delta$, $\tau$, $\rho_\zeta$, $\sigma_\zeta$, the transition probabilities for idiosyncratic employment states, and $\mu_\lambda$.

\paragraph{Equilibrium definition and computation.}
The mathematical definition of a recursive competitive equilibrium is standard, and we refer to \citet{Winberry2016} for details. We now review \citeauthor{Winberry2016}'s method for solving the model numerically.

A key model object is the cross-sectional joint distribution of the micro state variables, i.e., employment status $\epsilon_{i,t}$, normalized assets $a_{i,t-1}$, and permanent productivity $\lambda_i$. This distribution, which we denote $\tilde{\mu}_t(\epsilon,a,\lambda)$, is time-varying as it implicitly depends on the aggregate productivity state variable $\zeta_t$ at time $t$. Due to log utility and the linearity of the households' budget constraint in $\lambda_i$, macro aggregates are unaffected by the distribution of the permanent cross-sectional heterogeneity $\lambda_i$ (recall that $\mathbb{E}[\lambda_i]=1$). This implies that the mean parameter $\mu_\lambda$ of the log-normal distribution of $\lambda_i$ is only identifiable if micro data is available, as discussed further in \cref{sec:example_hh}. In equilibrium we have $\tilde{\mu}_t(\epsilon,a,\lambda) = \mu_t(\epsilon,a)F(\lambda \mid \mu_\lambda)$, where $F(\cdot \mid \mu_\lambda)$ denotes the time-invariant log-normal distribution for $\lambda_i$.

To solve the model numerically, \citet{Winberry2016,Winberry2018} assumes that the \emph{infinite-dimen\-sional} cross-sectional distribution $\mu_t(\epsilon,a)$ can be well approximated by a rich but \emph{finite-dimensional} family of distributions. The distribution of $a$ given $\epsilon$ is a mixture of a mass point at 0 (the borrowing constraint) and an absolutely continuous distribution concentrated on $(0,\infty)$. At every point in time, \citeauthor{Winberry2016} approximates the absolutely continuous part using a density of the exponential form \[g_{\epsilon}(a) = \exp\left\{\tilde\varphi_{\epsilon0}+\tilde\varphi_{\epsilon1}\tilde m_{\epsilon1}+\sum_{\ell=2}^q \varphi_{\epsilon\ell} \left[\left(a-\tilde m_{\epsilon1}\right)^\ell-\tilde m_{\epsilon\ell}\right]\right\},\] where $\tilde m_{\epsilon1}=\mathbb E[a \mid \epsilon]$, $\tilde m_{\epsilon l}=\mathbb E[\left(a-\tilde m_{\epsilon1}\right)^\ell\mid \epsilon]$ for $l\ge2$, the $\tilde\varphi_{\epsilon\ell}$'s are coefficients of the distribution, and $q \in \mathbb{N}$ is a tuning parameter that determines the quality of the numerical approximation.
The $q+1$ coefficients $\{\tilde\varphi_{\epsilon\ell}\}_{0\le l\le q}$ are pinned down by the $q$ moments $\{\tilde m_{\epsilon\ell}\}_{1\le l\le q}$, along with the normalization that $g_{\epsilon}(a)$ integrates to one. The approximation of the distribution $\mu_t(\epsilon,a)$ at any point in time therefore depends on $2(q+1)$ parameters: the probability point mass at $a=0$ as well as the $q$ moments $\{\tilde m_{\epsilon\ell}\}_{1\le l\le q}$, for each employment state $\epsilon$. Denote the vector of all these parameters by $\psi$. The model solution method proceeds under the assumption $\mu_t(a,\epsilon) = G(a,\epsilon; \psi_t)$, where $G$ denotes the previously specified parametric mixture functional form for the distribution, and we have added a time subscript to the parameter vector $\psi = \psi_t$. Though the approximation $\mu_t(a,\epsilon) \approx G(a,\epsilon; \psi_t)$ only becomes exact in the limit $q \to \infty$, the approximation may be good enough for small $q$ to satisfy the model's equilibrium equations to a high degree of numerical accuracy.

Adopting the distributional approximation, the model's aggregate equilibrium can now be written as a nonlinear system of expectational equations in a finite-dimensional vector $z_t$ of macro variables:
\begin{equation} \label{eqn:equilibrium_nonlin}
\mathbb{E}_t[H(z_{t+1}, z_t,\varepsilon_{t+1}; \theta)] = 0,
\end{equation}
where we have made explicit the dependence on the deep model parameters $\theta$. Consistent with the notation in \cref{sec:framework}, the vector $z_t$ includes (log) aggregate output $\log(Y_t)$, capital $\log(K_t)$, wages $\log(w_t)$, rate of return $r_t$, and productivity $\zeta_t$, but also the time-varying distributional parameters $\psi_t$. For brevity, we do not specify the full equilibrium correspondence $H(\cdot)$ here but refer to \citet{Winberry2016} for details. Among other things, $H(\cdot)$ enforces that the evolution over time of the cross-sectional distributional parameters $\psi_t$ is consistent with households' optimal savings decision rule, given the other macro state variables in $z_t$. $H(\cdot)$ also enforces consistency between micro variables and macro aggregates, such as capital market clearing $K_t = \sum_{\epsilon=0}^1 \int a \mu_t(\epsilon,da)$.

Estimation of the heterogenous agent model requires a fast numerical solution method, which \citet{Winberry2016,Winberry2018} achieves using the \citet{Reiter2009} linearization approach. First the system of equations \eqref{eqn:equilibrium_nonlin} is solved numerically for the steady state values $z_t=z_{t-1}=\bar{z}$ in the case of no aggregate shocks ($\varepsilon_t=0$). Then the system \eqref{eqn:equilibrium_nonlin} is linearized as a function of the aggregate variables $z_t$, $z_{t-1}$, and $\varepsilon_t$ around their steady state values, and the unique bounded rational expectations solution is computed (if it exists) using standard methods for linearized models \citep{Herbst2016}. This leads to a familiar linear transition equation of the form:
\begin{equation} \label{eqn:loglin_transition}
z_t - \bar{z} = A(\theta)(z_{t-1}-\bar{z}) + B(\theta)\varepsilon_t.
\end{equation}
The matrices $A(\theta)$ and $B(\theta)$ are functions of the derivatives of the equilibrium correspondence $H(\cdot)$, evaluated at the steady state $\bar{z}$. Notice that $A(\cdot)$ and $B(\cdot)$ implicitly depend on functionals of the steady-state cross-sectional  distribution of the micro state variables $(\epsilon,a)$.  This is because the \citet{Reiter2009} approach only linearizes with respect to \emph{macro} aggregates $z_t$ and shocks $\varepsilon_t$, while allowing for all kinds of heterogeneity and nonlinearities on the micro side, such as the borrowing constraint in the present model. In practice, \citet{Winberry2016,Winberry2018} implements the linearization of equation \eqref{eqn:equilibrium_nonlin} automatically through the software package Dynare.\footnote{See \citet{Adjemianetal2011}.} For pedagogical purposes, we build our inference machinery on top of the code that \citeauthor{Winberry2018} kindly makes available on his website, but we discuss alternative cutting-edge model solution methods in \cref{sec:concl}.

Our inference method treats the linearized equilibrium relationship \eqref{eqn:loglin_transition} as the true model for the (partially unobserved) macro aggregates $z_t$. That is, we do not attempt to correct for approximation errors due to linearization or due to the finite-dimensional approximation of the micro distribution. In particular, the transition density $p(z_t \mid z_{t-1},\theta)$ introduced in \cref{sec:framework} is obtained from the linear Gaussian dynamic equation \eqref{eqn:loglin_transition}, as opposed to the exact nonlinear equilibrium of the model, which is challenging to compute. We stress that the goal of our paper is to fully exploit \emph{all} observable implications of the (numerically approximated) structural model, and we leave concerns about model misspecification to future work (see also \cref{sec:concl}).
 
\paragraph{Sampling densities.}
We now show how the sampling densities of macro and micro data can be derived from the numerical model equilibrium.

For sake of illustration, assume that we observe a single macro variable given by a noisy measure of log output, i.e., $x_t=\log(Y_t)+e_t$, where $e_t \sim N(0,\sigma_e^2)$. The measurement error is not necessary for our method to work; we include it to illustrate the identification status of different kinds of parameters in \cref{sec:example_hh}. For this choice of observable, the sampling density $p(x_t \mid z_t,\theta)$ introduced in \cref{sec:framework} is given by a normal density with mean $\log(Y_t)$ and variance $\sigma_e^2$. More generally, we could consider a vector of macro observables $x_t$ linearly related to the state variables $z_t$, with a vector $e_t$ of additive measurement error:\footnote{Some of the elements of $e_t$ could have variance 0 if no measurement error is desired.}
\begin{equation} \label{eqn:loglin_meas}
x_t = S(\theta)z_t+e_t.
\end{equation}
Together, the equations \eqref{eqn:loglin_transition}--\eqref{eqn:loglin_meas} constitute a linear state space model in the observed and unobserved macro variables. We exploit this fact to evaluate the macro and micro likelihood function in \cref{sec:method}.

As for the micro data, suppose additionally that we observe repeated cross sections of households' employment status $\epsilon_{i,t}$ and after-tax/after-benefits income $\iota_{i,t} = \lambda_i\lbrace w_t[(1-\tau)\epsilon_{i,t} + b(1-\epsilon_{i,t})] + (1+r_t)a_{i,t-1} \rbrace$. That is, at certain times $t \in \mathcal{T} = \lbrace t_1,t_2,\dots,t_{|\mathcal{T}|} \rbrace$ we observe $N_t$ observations $y_{i,t} = (\epsilon_{i,t},\iota_{i,t})'$, $i=1,\dots,N_t$, drawn independently from a cross-sectional distribution that is consistent with $\mu_t(\epsilon,a)$, $F(\lambda \mid \mu_\lambda)$, and $z_t$. The joint sampling density $p(y_{i,t} \mid z_t,\theta)$ can be derived from the model's underlying cross-sectional distributions. The conditional distribution of $\iota_{i,t}$ given $\epsilon_{i,t}$ and the macro states is absolutely continuous, since the micro heterogeneity $\lambda_i$ smooths out the point mass at the households' borrowing constraint. By differentiating the cumulative distribution function, it can be verified that the conditional sampling density of $\iota_{i,t}$ given $\epsilon_{i,t}$ equals
\begin{equation} \label{eqn:hh_income_dens}
p(\iota_{i,t} \mid \epsilon_{i,t},z_t,\theta) = \pi_{\epsilon_{i,t}}(\psi_t) \frac{f(\frac{\iota_{i,t}}{\xi_{i,t}} \mid \mu_{\lambda})}{\xi_{i,t}} + [1-\pi_{\epsilon_{i,t}}(\psi_t)] \int_{0}^\infty \frac{f(\frac{\iota_{i,t}}{\xi_{i,t}+(1+r_t) a} \mid \mu_{\lambda})}{\xi_{i,t}+(1+r_t) a}g_{\epsilon_{i,t}}(a \mid \psi_t)\,da,
\end{equation}
where $f(\cdot \mid \mu_\lambda)$ is the assumed log-normal density for $\lambda_i$, $\pi_\epsilon(\psi_t) \equiv P(a=0 \mid \epsilon,\psi_t)$ is the probability mass at zero for assets, and $\xi_{i,t} \equiv w_t[(1-\tau)\epsilon_{i,t} + b(1-\epsilon_{i,t})]$. In practice, the integral can be evaluated numerically, cf.\ \cref{sec:example_hh}.

This concludes the specification of the model as well as the derivations of the macro state transition density and of the sampling densities for the macro and micro data. In \cref{sec:method} we will use these ingredients to derive the likelihood function consistent with the model and the observed data.

\paragraph{Other observables and models.}
Of course, one could think of many other empirically relevant choices of macro and micro observables, leading to other expressions for the sampling densities. Our choices here are merely meant to illustrate how our framework is flexible enough to accommodate: (i) a mixture of discrete and continuous observables; (ii) observables that depend on both micro and macro states; and (iii) persistent cross-sectional heterogeneity $\lambda_i$ that, given repeated cross section data, effectively amounts to measurement error at the micro level.

We emphasize that the general framework in \cref{sec:framework} can also handle many other types of heterogeneous agent models. To show this, \cref{sec:example_firm} will consider an alternative model with heterogeneous firms as in \citet{Khan2008}.

\section{Efficient Bayesian inference}
\label{sec:method}
We now describe our method for doing efficient Bayesian inference. We first construct a numerically unbiased
estimate of the likelihood, and then discuss the posterior sampling procedure. Finally, we compare our approach with procedures that collapse the micro data to a set of cross-sectional moments.

\subsection{Unbiased likelihood estimate}
Our likelihood estimate is based on decomposing the joint likelihood into a macro part and a micro part (conditional on the macro data):
\begin{align}
p(\mathbf{x},\mathbf{y} \mid \theta) &= \overbrace{p(\mathbf{x} \mid \theta)}^{\text{macro}} \, \overbrace{p(\mathbf{y} \mid \mathbf{x},\theta)}^{\text{micro}} \nonumber \\
&= p(\mathbf{x} \mid \theta) \int p(\mathbf{y} \mid \mathbf{z}, \theta) p(\mathbf{z} \mid \mathbf{x},\theta)\, d\mathbf{z} \nonumber
\\
&= p(\mathbf{x} \mid \theta) \int \prod_{t \in \mathcal{T}}\prod_{i=1}^{N_t} p(y_{i,t} \mid z_t,\theta) p(\mathbf{z} \mid \mathbf{x},\theta) \, d\mathbf{z}. \label{eqn:micro_likel_int}
\end{align}
Note that this decomposition is satisfied by construction under \cref{asn:model} and will purely serve as a computational tool. The form of the decomposition should not be taken to mean that we are assuming that ``$\mathbf{x}$ affects $\mathbf{y}$ but not \emph{vice versa}.'' As discussed in \cref{sec:model}, our framework allows for a fully general equilibrium feedback loop between macro and micro variables.

The macro part of the likelihood is easily computable from the \citeauthor{Reiter2009}-linearized state space model \eqref{eqn:loglin_transition}--\eqref{eqn:loglin_meas}. Assuming i.i.d.\ Gaussian measurement error $e_t$ and macro shocks $\varepsilon_t$, the macro part of the likelihood $p(\mathbf{x} \mid \theta)$ can be obtained from the Kalman filter. This is computationally cheap even in models with many state variables and/or observables. This idea was developed by \citet{Mongey2017} and \citet{Winberry2018} for estimation of heterogeneous agent models from aggregate time series data.

The novelty of our approach is that we compute an unbiased estimate of the micro likelihood conditional on the macro data. Although the integral in expression \eqref{eqn:micro_likel_int} cannot be computed analytically in realistic models, we can obtain a \emph{numerically unbiased} estimate of the integral by random sampling:
\begin{equation} \label{eqn:micro_likel_unbias}
\int \prod_{t \in \mathcal{T}}\prod_{i=1}^{N_t} p(y_{i,t} \mid z_t,\theta) p(\mathbf{z} \mid \mathbf{x},\theta) \, d\mathbf{z} \approx \frac 1 J \sum_{j=1}^J \prod_{t \in \mathcal{T}}\prod_{i=1}^{N_t} p(y_{i,t} \mid z_t=z_t^{(j)},\theta),
\end{equation}
where $\mathbf{z}^{(j)} \equiv \lbrace z_t^{(j)} \rbrace_{1 \leq t \leq T}$, $j=1,\dots,J$, are draws from the joint smoothing density $p(\mathbf{z} \mid \mathbf{x},\theta)$ of the latent states. Again using the \citeauthor{Reiter2009}-linearized model solution, the Kalman smoother can be used to produce these state smoothing draws with little computational effort \citep[e.g.,][]{Durbin2002}. As the number of smoothing draws $J \to \infty$, the likelihood estimate converges to the exact likelihood, but we show below that finite $J$ is sufficient for our purposes, as we rely only on the numerical unbiasedness of the likelihood estimate, not its consistency.

Our likelihood estimate can loosely be interpreted as arising from a two-step approach: First we estimate the states from the macro data, and then we plug the state estimates $z_t^{(j)}$ into the micro sampling density. However, unlike more \emph{ad hoc} versions of this general idea, we will argue next that the unbiased likelihood estimate makes it possible to perform valid Bayesian inference that fully takes into account all sources of uncertainty about states and parameters.

The expression on the right-hand side of the likelihood estimate \eqref{eqn:micro_likel_unbias} is parallelizable over smoothing draws $j$, time $t$, and/or individuals $i$. Thus, given the right computing environment, the computation time of our method scales well with the dimensions of the micro data. This is particularly helpful in models where evaluation of the micro sampling density involves numerical integration, as in the household model in \cref{sec:example_hh} below.

\subsection{Posterior sampling}
Now that we have a numerically unbiased estimate of the likelihood, we can plug it into any generic MCMC procedure to obtain draws from the posterior distribution, given a choice of prior density. We may simply pretend that the likelihood estimate is exact and run the MCMC algorithm as we otherwise would, as explained by \citet{Andrieu2010} and \citet{Flury2011}. Despite the simulation error in estimating the likelihood, the ergodic distribution of the MCMC chain will equal the fully efficient posterior distribution $p(\theta \mid \mathbf{x},\mathbf{y})$. This is true no matter how small the number $J$ of smoothing draws is. Still, the MCMC chain will typically exhibit better mixing if $J$ is moderately large so that proposal draws are not frequently rejected merely due to numerical noise. In principle, we can use any generic MCMC method that requires only the likelihood and prior density as inputs, such as Metropolis-Hastings. Our approach can also be applied to Sequential Monte Carlo sampling \citep[chapter 5]{Herbst2016}.\footnote{Implementation of Algorithm 8 in \citet{Herbst2016} requires some care. The mutation step (step 2.c) can use the unbiased likelihood estimate with finite number of smoothing draws $J$, by \citet{Andrieu2010}. However, it is not immediately clear whether an unbiased likelihood estimate suffices for the correction step (step 2.a). For the latter step, we therefore advise using a larger number of smoothing draws to ensure that the likelihood estimate is close to its analytical counterpart.}

\subsection{Comparison with moment-based methods}
\label{sec:theory_moment}

The above full-information approach yields draws from the same posterior distribution as if we had used the model-implied exact joint likelihood of the micro and macro data; it is thus finite-sample optimal in the usual sense. An alternative approach proposed by \citet{Challe2017} and \citet{Mongey2017} is to collapse the micro data into a small number of cross-sectional moments which are tracked over time (that is, the repeated cross sections of micro data are transformed into time series of cross-sectional moments). We now examine under which circumstances our full-information approach is strictly more efficient than this moment-based approach.

We focus on moment-based approaches that track the evolution of cross-sectional moments over time, rather than exploiting only steady-state moments. Empirically, cross-sectional distributions are often time-varying \citep{krueger2010cross,wolff2016household}. The model-based numerical illustrations below also exhibit time-variation in cross-sectional distributions. Thus, collapsing the time-varying moments to averages across the entire time sample would leave information on the table.

If the micro sampling density has sufficient statistics for the parameters of interest, and the sufficient statistics are one-to-one functions of the observed cross-sectional moments, then these moments contain the same amount of information about the structural parameters as the full micro data set. As stated in the Pitman-Koopman-Darmois theorem, only the exponential family has a fixed number of sufficient statistics. The following result obtains.

\begin{thm} \label{thm:suff_stat}
If the conditional sampling density of the micro data $y_{i,t}$ can be expressed as
\begin{equation} \label{eqn:exp}
p(y_{i,t} \mid z_t,\theta)
=\exp\left[\varphi_0(z_t,\theta)+\mathfrak m_{0}(y_{i,t})+\sum_{\ell=1}^{Q} \varphi_\ell(z_t,\theta)\mathfrak m_{\ell}(y_{i,t})\right],
\end{equation}
for certain functions $\varphi_\ell(\cdot,\cdot),\mathfrak m_\ell(\cdot)$, $\ell=0,\dots,Q$, then there exist sufficient statistics for $\theta$ given by the cross-sectional moments
\begin{equation}
\hat m_{\ell,t} = \frac 1{N_t} \sum_{i=1}^{N_t} \mathfrak m_{\ell}(y_{i,t}),\quad \ell=1,\dots,Q. \label{eqn:moments}
\end{equation}
\end{thm}
\begin{proof}
Please see \cref{sec:moment_proof}.
\end{proof}
That is, under the conditions of the theorem, the full micro-macro data set $\lbrace \mathbf{y},\mathbf{x} \rbrace$ contains as much information about the parameters $\theta$ as the moment-based data set $\lbrace \mathbf{\hat{m}},\mathbf{x} \rbrace$, where $\mathbf{\hat{m}} \equiv \lbrace \hat{m}_{\ell,t} \rbrace_{1 \leq \ell \leq Q, t \in \mathcal{T}}$. This result is not trivial due to the presence of the latent macro states $z_t$, which are integrated out in the likelihood \eqref{eqn:micro_likel_int}. The key requirement is that in \eqref{eqn:exp}, the terms inside the exponential should be additive and each term should take the form \(\varphi_\ell(z_t,\theta)\mathfrak m_\ell(y_{i,t})\).

Whether or not the micro sampling density exhibits the exponential form \eqref{eqn:exp} depends on the model and on the choice of micro observables. As explained in \cref{sec:model_hh}, in this paper we adopt the \citet{Winberry2018} model solution approach, which approximates the cross-sectional distribution of the idiosyncratic micro state variables $s_{i,t}$ using an exponential family of distributions. Hence, \emph{if we observed the micro states $s_{i,t}$ directly}, \cref{thm:suff_stat} implies that there would be no loss in collapsing the micro data to a certain set of cross-sectional moments. However, there may not exists sufficient statistics for the actual micro observables $y_{i,t}$, which are generally non-trivial functions of the latent micro states $s_{i,t}$ and macro states $z_t$. The following corollary gives conditions under which sufficient statistics still obtain. Let \(y_{i,t}\) be a \(d_y\times1\) vector and \(s_{i,t}\) be a \(d_s\times1\) vector.

\begin{cor} \label{cor:exp_poly}
Suppose we have:
\begin{enumerate}
\item The conditional density of the micro states $s_{i,t}$ given $z_t$ is of the exponential form.
\item The micro states are related to the micro observables as follows:
\begin{equation}
s_{i,t}=B_1(z_t,\theta)\Upsilon(y_{i,t})+B_0(z_t,\theta), \label{eqn:transform}
\end{equation}
where:
\begin{enumerate}[a)]
\item \(d_{y}=d_{s}\).
\item $\Upsilon(\cdot)$ is a known, piecewise bijective and differentiable function with its domain and range being subsets of \(\mathbb R^{d_s}\).
\item The $d_s \times d_s$ matrix $B_1(z_t,\theta)$ is non-singular for almost all values of \((z_t,\theta)\). 
\end{enumerate}
\end{enumerate}
Then there exist sufficient micro statistics for $\theta$.
\end{cor}
\begin{proof}
Please see \cref{sec:moment_proof}. The proof states the functional form of the sufficient statistics.
\end{proof}
There are several relevant cases where the conditions of \cref{cor:exp_poly} fail and hence sufficient statistics may not exist. First, the dimension $d_y$ of the observables $y_{i,t}$ could be strictly smaller than the dimension $d_s$ of the latent micro states $s_{i,t}$. Second, there may not exist any linear relationship between $s_{i,t}$ and some function $\Upsilon(y_{i,t})$ of $y_{i,t}$, for example due to binding constraints. Third, there may be unobserved individual heterogeneity and/or micro measurement error, such as the individual-specific productivity parameter $\lambda_i$ in \cref{sec:model_hh}. We provide further discussion in \cref{sec:suff_stat_general}.

Even if the model exhibits sufficient statistics given by cross-sectional moments of the observed micro data, valid inference requires taking into account the sampling uncertainty of these moments. This is a challenging task, since the finite-sample distribution of the moments is typically not Gaussian (especially for higher moments), see \cref{sec:nonlinear-Gaussian} for an example. Hence, the observation equation for the moments does not fit into the linear-Gaussian state space framework \eqref{eqn:loglin_transition}--\eqref{eqn:loglin_meas} that lends itself to Kalman filtering. If the micro sample size is large, the sampling distribution of the moments may be well approximated by a Gaussian distribution, but even then the variance-covariance matrix of the distribution will generally be time-varying and difficult to compute/estimate. In \cref{sec:example_hh} below we consider one natural method for approximately accounting for the sampling uncertainty of the moments. We find that this moment-based approach is less reliable than our preferred full-information approach.

The potential inefficiency and fragility of the moment-based approach contrasts with the ease of applying our efficient full-information method. Users of our method need not worry about the existence of sufficient statistics, nor do they need to select which moments to include in the analysis and figure out how to account for their sampling uncertainty. Moreover, we show by example in \cref{sec:example_firm_censor} that the full-information approach can easily accommodate empirically relevant features of micro data such as censoring or selection, which is challenging to do in a moment-based framework (at least in an efficient way).

\section{Illustration: Heterogeneous household model}
\label{sec:example_hh}

We now demonstrate that combining macro and micro data can sharpen structural inference when estimating the heterogeneous household model of \cref{sec:model_hh} on simulated data. We contrast the results of our efficient full-information approach with those of an alternative moment-based approach. This section should be viewed as a proof-of-concept exercise, as we deliberately keep the dimensionality of the inference problem small in order to focus attention on the core workings of our procedure.

\subsection{Model, data, and prior}
We consider the stylized heterogeneous household model defined in \cref{sec:model_hh}. We aim to estimate the households' discount factor $\beta$, the standard deviation $\sigma_e$ of the measurement error in log output, and the individual productivity heterogeneity parameter $\mu_\lambda$. All other parameters are assumed known for simplicity.

Consistent with \cref{sec:model_hh}, we assume that the econometrician observes aggregate data on log output with measurement error, as well as repeated cross sections of household employment status $\epsilon_{i,t}$ and after-tax/after-benefits income $\iota_{i,t}$.

We adopt the annual parameter calibration in \citet{Winberry2016}, see \cref{sec:example_hh_calib}. In particular, $\beta=0.96$. We choose the true measurement error standard deviation $\sigma_e$ so that about 20\% of the variance of observed log output is due to measurement error, yielding $\sigma_e=0.02$.\footnote{One possible real-world interpretation of the measurement error is that it represents the statistical uncertainty in estimating the natural rate of output (recall that the model abstracts from nominal rigidities).} The individual heterogeneity parameter $\mu_\lambda$ is chosen to be $-0.25$, implying that the model's cross-sectional 20th to 90th percentile range of log after-tax income roughly matches the range in U.S. data \citep[Table I]{Piketty2018}.

Using this calibration, we simulate $T=100$ periods of macro data, as well as micro data consisting of $N_t=N=1,000$ households observed at each of the ten time points $t=10,20,30,\dots,100$. The data is simulated using the same approximate model solution method as is used to compute the unbiased likelihood estimate, see \cref{sec:model_hh}.

Finally, we choose the prior on $(\beta,\sigma_e,\mu_\lambda)$ to be flat in the natural parameter space.

\subsection{Computation}
Following \citet{Winberry2016,Winberry2018}, we solve the model using a Dynare implementation of the \citet{Reiter2009} method. This allows us to use Dynare's built-in Kalman filter/smoother procedures when evaluating the micro likelihood estimate \eqref{eqn:micro_likel_unbias}. We use an approximation of degree $q=3$ when approximating the asset distribution, in the notation of \cref{sec:model_hh}. We average the likelihood across $J=500$ smoothing draws. The integral \eqref{eqn:hh_income_dens} in the micro sampling density of income is evaluated using a combination of numerical integration and interpolation.\footnote{First, we use a univariate numerical integration routine to evaluate the integral on an equal-spaced grid of values for $\log \iota$. Then we use cubic spline interpolation to evaluate the integral at arbitrary $\iota$. In practice, a small number of grid points is sufficient in this application, since the density \eqref{eqn:hh_income_dens} is a smooth function of $\iota$.} To simulate micro data from the cross-sectional distribution, we apply the inverse probability transform to the model-implied cumulative distribution function of assets, which in turn is computed using numerical integration.

For simplicity, our MCMC algorithm is a basic Random Walk Metropolis-Hastings algorithm with tuned proposal covariance matrix and adaptive step size \citep{Atchade2005}.\footnote{Our proposal distribution is a mixture of (i) the adapted multivariate normal distribution and (ii) a diffuse normal distribution, with 95\% probability attached to the former. We verified the Diminishing Adaption condition and Containment condition in \citet{rosenthal2011optimal}, so the distribution of the MCMC draws will converge to the posterior distribution of the parameters.} The starting values are determined by a rough grid search on the simulated data. We generate 10,000 draws and discard the first 1,000 as burn-in. Using parallel computing on 20 cores, likelihood evaluation takes about as long as \citeauthor{Winberry2016}'s (\citeyear{Winberry2016}) procedure for computing the model's steady state.

\subsection{Results}
\label{sec:example_hh_post}
\begin{figure}[t]
\centering
\textsc{Heterogeneous household model: Posterior density} \\[\baselineskip]
\includegraphics[width=\linewidth]{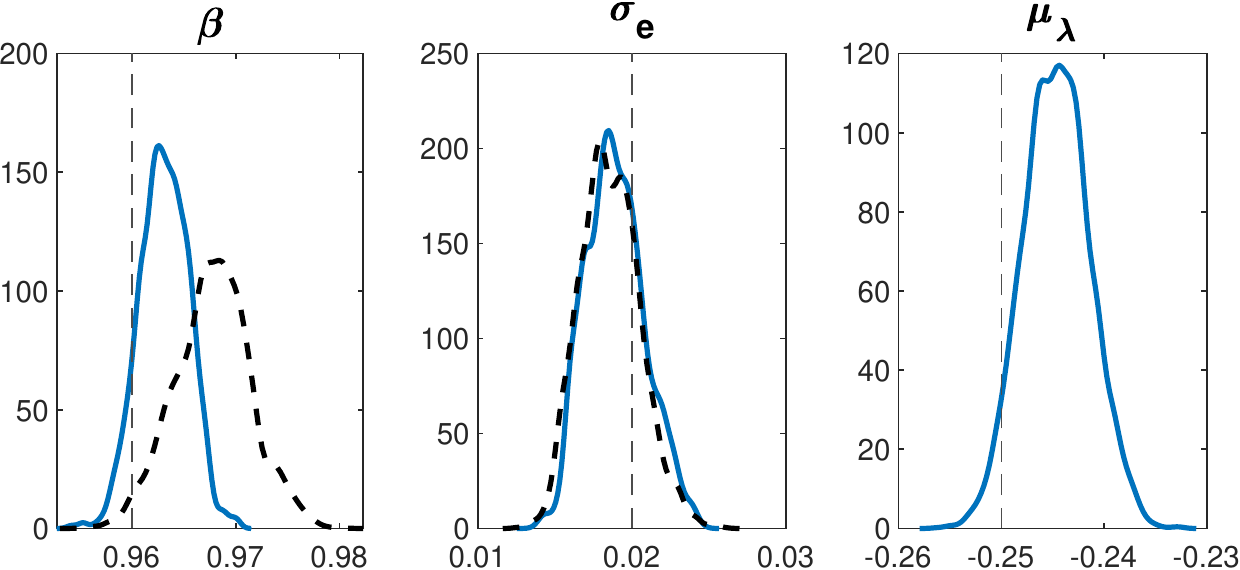}
\caption{Posterior densities with (blue solid curves) and without (black dashed curves) conditioning on the micro data. Both sets of results use the same simulated data set. Vertical dashed lines indicate true parameter values. Posterior density estimates from the 9,000 retained MCMC draws using Matlab's {\tt ksdensity} function with default settings. The third display omits the macro-only results, since $\mu_\lambda$ is not identified from macro data alone.}
\label{fig:example_hh_postdens}
\end{figure}

\cref{fig:example_hh_postdens} shows that both macro and micro data can be useful or even essential for estimating some parameters, but not others. The figure depicts the posterior densities of the three parameters, on a single sample of simulated data. The full-information posterior (blue solid curves) is concentrated close to the true values of the three parameters (which are marked by vertical thin dashed lines). The figure also shows the posterior density \emph{without} conditioning on the micro data (black dashed curves). The household discount factor $\beta$ is an important determinant of not just aggregate variables, but also the heterogeneous actions of the micro agents in the economy. Ignoring the micro data leads to substantially less accurate inference about $\beta$ in this simulation, as the macro-only posterior is less precisely centered around the true value as well as more diffuse than the full-information posterior. Nevertheless, macro data clearly does meaningfully contribute to pinning down the parameter $\beta$.
More starkly, $\mu_\lambda$ can only be identified from the cross section, since by construction the macro aggregates are not influenced by the distribution of the individual permanent productivity draws $\lambda_i$. In contrast, essentially all the information about the measurement error standard deviation $\sigma_e$ comes from the macro data, again by construction. Thus, our results here illustrate the general lesson that both macro and micro data can be either essential, useful, or irrelevant for estimating different parameters.

\begin{figure}[t]
\centering
\textsc{Heterogeneous household model: Consumption policy function, employed} \\[\baselineskip]
\includegraphics[width=\linewidth,clip=true,trim=2em 0 2em 0]{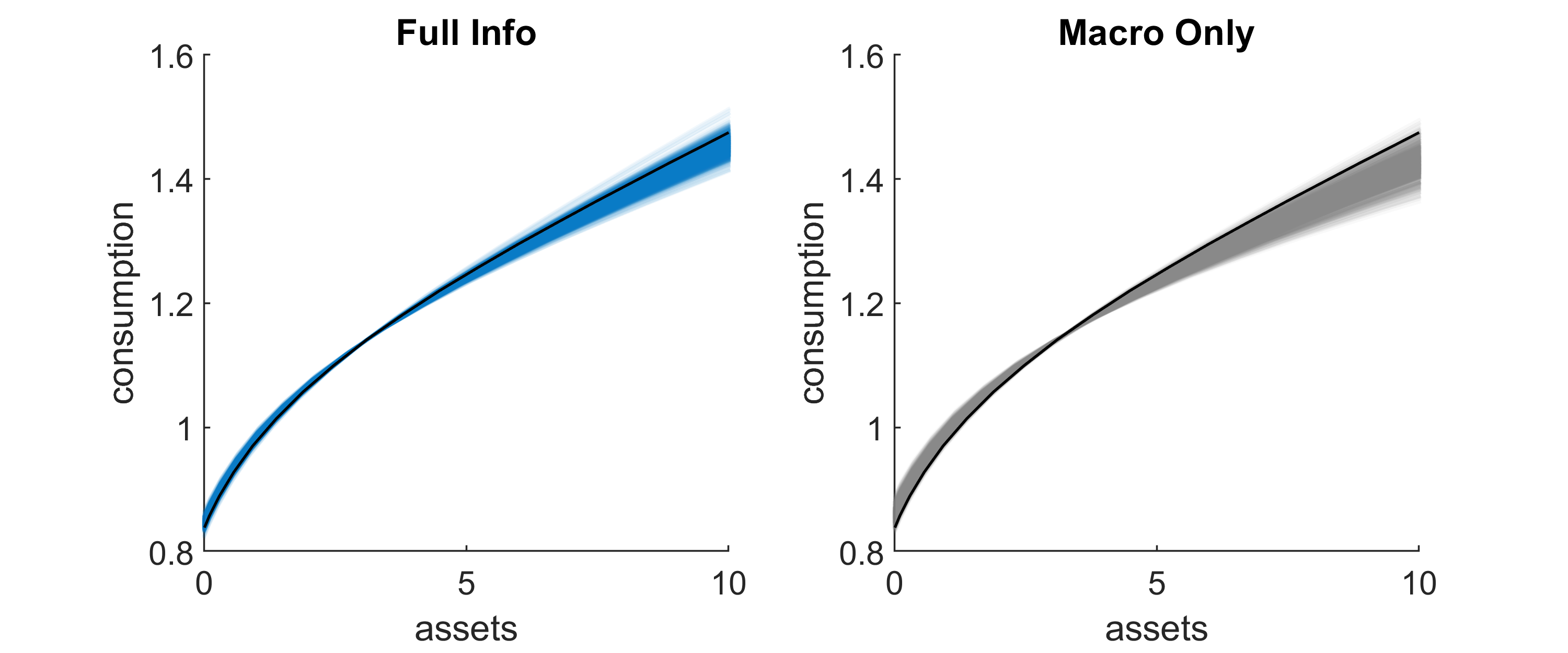} 
\caption{Estimated steady state consumption policy function for employed households, either using both micro and macro data (left panel) or only using macro data (right panel). The thick black curve is computed under the true parameters. The thin lines are 900 posterior draws (computed using every 10th MCMC draw after burn-in). X-axes are normalized asset holdings $a_{i,t}$.}
\label{fig:example_hh_consumption}
\end{figure}

\cref{fig:example_hh_consumption} shows that efficient use of the micro data leads to substantially more precise estimates of the steady state consumption policy function for employed households.\footnote{\cref{fig:example_hh_consumption_unemp} in \cref{sec:example_hh_results} plots the policy function for unemployed households.} The left panel shows that full-information posterior draws of the consumption policy function (thin curves) are fairly well centered around the true function (thick curve), as is expected given the accurate inference about $\beta$ depicted in \cref{fig:example_hh_postdens}. In contrast, the right panel shows that macro-only posterior draws are less well centered and exhibit higher variance, especially for households with high or low current asset holdings. The added precision afforded by efficient use of the micro data translates into more precise estimates of the marginal propensity to consume (the derivative of the consumption policy function) at the extremes of the asset distribution. This is potentially useful when analyzing the two-way feedback effect between macroeconomic policies and redistribution \citep{Auclert2019}.

\begin{figure}[t]
\centering
\textsc{Het.\ household model: Impulse responses of asset distribution, employed} \\[\baselineskip]
\includegraphics[width=\linewidth,clip=true,trim=3em 0 2em 0]{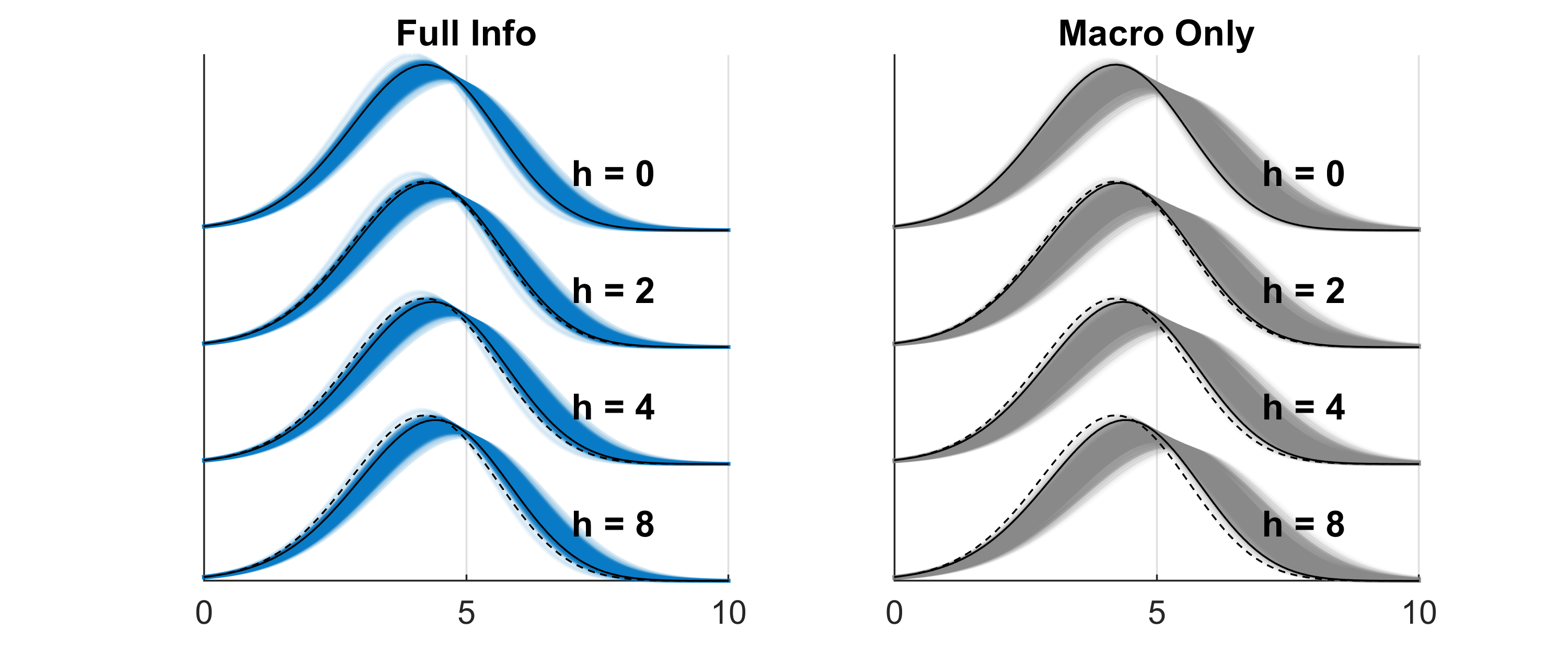} 
\caption{Estimated impulse response function of employed households' asset distribution with respect to an aggregate productivity shock, either using both micro and macro data (left panel) or only using macro data (right panel). The thin lines are 900 posterior draws (computed using every 10th MCMC draw after burn-in). X-axes are normalized asset holdings $a_{i,t}$. The four rows in each panel are the asset densities at impulse response horizons 0 (impact), 2, 4, and 8. The black dashed and black solid curves are the steady state density and the impulse response, respectively, computed under the true parameters. On impact the true impulse response equals the steady state density, since households' portfolio choice is predetermined.} 
\label{fig:example_hh_irf}
\end{figure}

The extra precision afforded by micro data also sharpens inference on the impulse response function of the asset distribution with respect to an aggregate productivity shock. \cref{fig:example_hh_irf} shows full-information (left panel) and macro-only  (right panel) posterior draws of the impulse response function of employed households' asset holding density, in the periods following a 5\%\ aggregate productivity shock.\footnote{For unemployed households, see \cref{fig:example_hh_irf_unemp} in \cref{sec:example_hh_results}.} Once again, the full-information results have substantially lower variance.  Following the shock, there is a noticeable movement of the asset distribution computed under the true parameters (black solid curve). At horizon $h=8$, the mean increases by 0.16 relative to the steady state (black dashed curve), the variance increases by 0.10, and the third central moment decreases by 0.06. However, the true movement in the asset distribution is not so large relative to the estimation uncertainty. This further motivates the use of an efficient inference method that validly takes into account all estimation uncertainty.

\begin{figure}[t]
\centering
\textsc{Heterogeneous household model: Posterior density, multiple simulations} \\[\baselineskip]
\includegraphics[width=\linewidth,clip=true,trim=3em 0 3em 0]{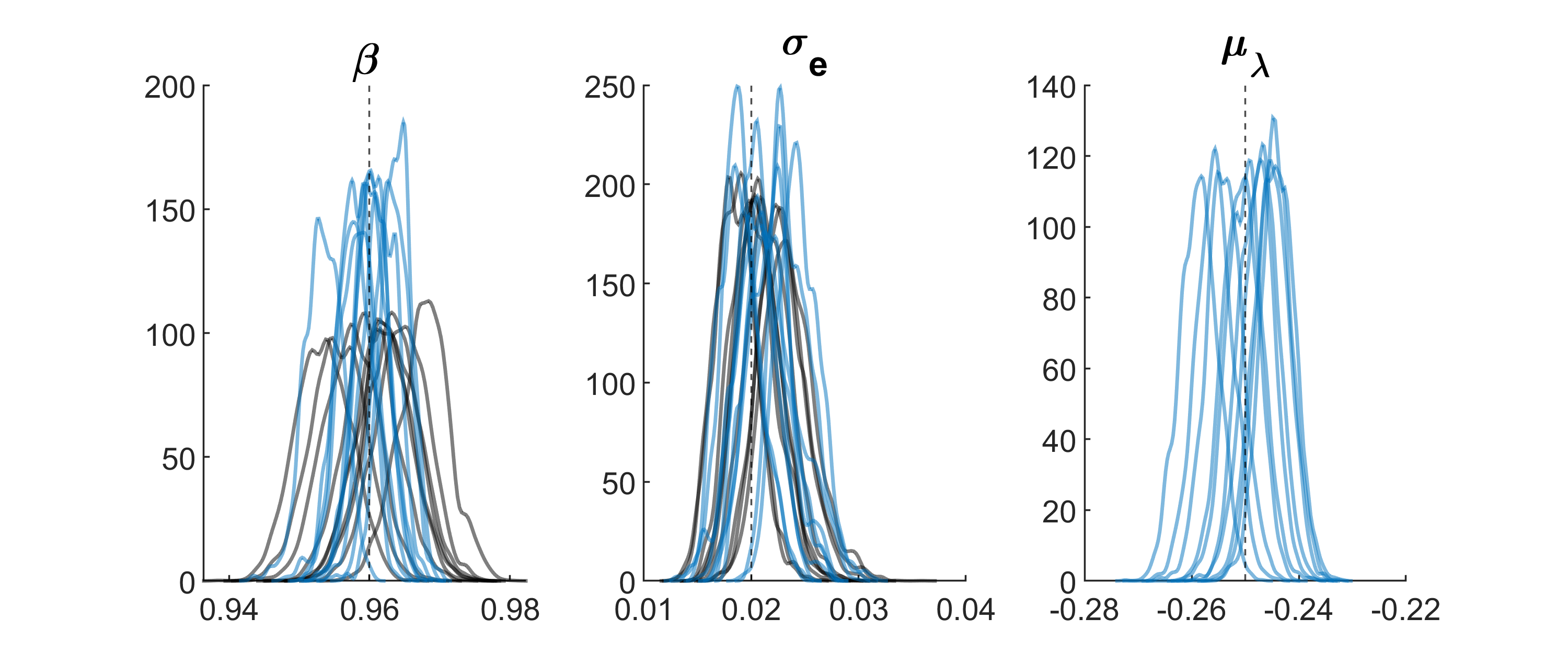}
\caption{Posterior densities with (blue curves) and without (gray curves) conditioning on the micro data, for 10 different simulated data sets. See also caption for \cref{fig:example_hh_postdens}.}
\label{fig:example_hh_postdens_all}
\end{figure}

The previous qualitative conclusions hold up in repeated simulations from the calibrated model. We repeat the MCMC estimation exercise on 10 different simulated data sets.\footnote{Computational constraints preclude a full Monte Carlo study.} \cref{fig:example_hh_postdens_all} plots all 10 full-information and macro-only posterior densities for the three parameters on the same plot. Notice that the full-information densities for $\beta$ systematically concentrate closer to the true value than the macro-only posteriors do, as in \cref{fig:example_hh_postdens}.

Our inference approach is valid in the usual Bayesian sense no matter how small the sample size is. In \cref{fig:example_hh_postdens_N100} of \cref{sec:example_hh_results} we show that the full-information approach still yields useful inference about the model parameters if we only observe $N=100$ observations every ten periods (instead of $N=1000$ as above).

\subsection{Comparison with moment-based methods}
\label{sec:example_hh_lik}
In this subsection, we compare the above full-information results with a moment-based inference approach, to shed light on the theoretical comparison in \cref{sec:theory_moment}. Due to the unobserved individual heterogeneity parameter $\lambda_i$, fixed-dimensional sufficient statistics do not exist in this model with the given observables.\footnote{The unobserved individual heterogeneity is observationally equivalent to micro measurement error given repeated cross sections of micro data.} Hence, we follow empirical practice and compute an \emph{ad hoc} selection of cross-sectional moments, including the sample mean, variance, and third central moment of household after-tax income. We compute the moments separately for the groups of employed and unemployed households, in each period $t=10,20,\dots,100$ where micro data is observed. We consider three moment-based approaches with different numbers of observables: The ``1st Moment'' approach only incorporates time series of sample means, the ``2nd Moment'' approach incorporates both sample means and variances, and the ``3rd Moment'' approach incorporates sample moments up to the third order.

Once we compute the time series of cross-sectional moments on the simulated data, we treat them as additional time series observables and proceed as in the ``Macro Only'' approach considered earlier. To account for the sampling uncertainty in the cross-sectional moments, we appeal to a Central Limit Theorem and treat the moments as jointly Gaussian, which is equivalent to adding measurement error in those state space equations that correspond to the moments. A natural and practical way to construct the variance-covariance matrix of the measurement error is to estimate its elements using higher-order sample moments of micro data. The variance-covariance matrix is actually time-varying according to the structural model, but since this would be challenging to account for, we treat it as fixed over the sample.\footnote{Higher-order sample moments are less accurate approximations to their population counterparts. Given an empirically relevant cross-sectional sample size, the resulting variance-covariance matrix would be even more imprecise if inferred period by period.} \cref{sec:moment_vcv} provides the details of how we estimate the variance-covariance matrix. The computation time of the moment-based likelihood functions is not much faster than our full-information approach, since the evaluation of the micro likelihood (which is specific to the full-information method) takes approximately the same amount of time as the calculation of the model's steady state (which is common to all methods), when implemented on a research cluster with 20 parallel workers.

\begin{figure}[t]
\centering
\textsc{Heterogeneous household model: Likelihood comparison} \\[\baselineskip]
\includegraphics[width=\linewidth]{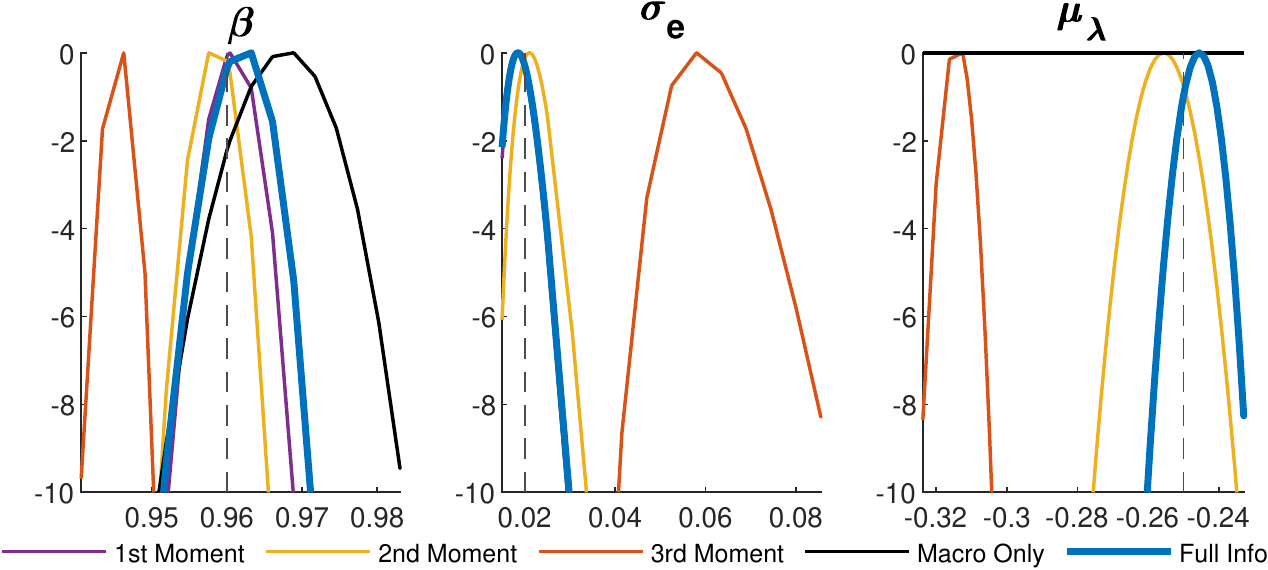}
\caption{Comparison of log likelihoods across inference methods, based on one typical simulated data set. Each panel depicts univariate deviations of a single parameter while keeping all other parameters at their true values. The maximum of each likelihood curve is normalized to be zero. Vertical dashed lines indicate true parameter values. The ``1st Moment'' and ``Macro Only'' curves are flat on the right panel, since $\mu_\lambda$ is not identified from this data alone. For results across 10 different simulated data sets, see \cref{fig:example_hh_lik_byest} in \cref{sec:example_hh_lik_multi}. }
\label{fig:example_hh_lik_byrep}
\end{figure}

We compare the shape and location of the likelihood functions for the full-information and moment-based methods.\footnote{We omit full posterior inference results for the moment-based methods, as they were more prone to MCMC convergence issues than our full-information method.} For graphical clarity, we vary a single parameter at a time, keeping the other parameters at their true values. \cref{fig:example_hh_lik_byrep} plots the univariate log likelihood functions of all inference approaches based on one typical simulated data set.\footnote{We compute the full-information likelihood function by averaging across $J=500$ smoothing draws. For a clearer comparison of the plotted likelihood functions, we fix the random numbers used to draw from the smoothing distribution across parameter values. Note that we \emph{do not} fix these random numbers in the MCMC algorithm, as required by the \citet{Andrieu2010} argument.} Since we are interested in the curvature of the likelihood functions near their maxima, and not the overall level of the functions, we normalize each curve by subtracting its maximum.

\cref{fig:example_hh_lik_byrep} shows that the moment-based likelihoods do not approximate the efficient full-information likelihood well, with the ``3rd Moment'' likelihood being particularly inaccurately centered. There are two reasons for this. First, as discussed in \cref{sec:theory_moment}, there is no theoretical sufficient statistics in this setup, so all the moment-based approaches incur some efficiency loss. Second, the sampling distributions of higher-order sample moments are not well approximated by Gaussian distributions in finite samples, and the measurement error variance-covariance matrix depends on even higher-order moments, which are poorly estimated. A separate issue is that the individual heterogeneity parameter $\mu_\lambda$ cannot even be identified using the ``1st Moment'' approach, since this parameter does not influence first moments of the micro data. The ``2nd Moment'' likelihood is not entirely misleading but nevertheless differs meaningfully from the full-information likelihood.\footnote{The ``Full Info'' and ``Macro Only'' likelihoods are consistent with the posterior densities plotted in \cref{sec:example_hh_post}. For $\beta$, the ``Macro Only'' likelihood has a smaller curvature around the peak and a wider range of peaks across simulated data sets, so the full information method helps sharpen the inference of $\beta$. For $\sigma_e$, the ``Macro Only'' curves are close to their ``Full Info'' counterparts. The parameter $\mu_\lambda$ is not identified in the ``Macro Only'' case, so the corresponding likelihood function is flat.} \cref{fig:example_hh_lik_byest} in \cref{sec:example_hh_lik_multi} confirms that the aforementioned qualitative conclusions hold up across 10 different simulated data sets.

To summarize, even in this relatively simple model, the moment-based methods we consider lead to a poor approximation of the full-information likelihood, and the inference can be highly sensitive to the choice of which moments to include. It is possible that other implementations of the moment-based approach would work better in particular applications. Nevertheless, any moment-based approach will require challenging \emph{ad hoc} choices, such as which moments to use and how to account for their sampling uncertainty. No such choices are required by the efficient full-information approach developed in this paper.

\section{Illustration: Heterogeneous firm model}
\label{sec:example_firm}
As our second proof-of-concept example, we estimate a version of the heterogeneous firm model of \citet{Khan2008}. In addition to showing that our general inference approach can be applied outside the specific \citet{Krusell1998} family of models, we use this section to illustrate how sample selection or data censoring can easily be accommodated in our method.

\subsection{Model, data, and prior}
\label{sec:example_firm_baseline}

A continuum of heterogeneous firms are subject to both idiosyncratic and aggregate productivity shocks. Investment is subject to non-convex adjustment costs. Specifically, firm $i$'s investment $I_{i,t}$ is free if $|I_{i,t}/k_{i,t}| \leq a$, where $k_{i,t}$ is the firm-specific capital stock, and $a \geq 0$ is a parameter. Otherwise, firms pay a fixed adjustment cost of $\xi_{i,t}$ in units of labor. $\xi_{i,t}$ is drawn at the beginning of every period from a uniform distribution on the interval $[0, \bar{\xi}]$, independently across firms and time. Here $\bar{\xi} \geq 0$ is another parameter. In addition to the aggregate productivity shock, there is a second aggregate shock that affects investment efficiency. The representative household has additively separable preferences over log consumption and (close to linear) leisure time. For brevity, we relegate the details of the model to \cref{sec:example_firm_details}, which entirely follows \citeauthor{Winberry2018}'s (\citeyear{Winberry2018}) version of the \citet{Khan2008} model.

We aim to estimate the adjustment cost parameters $\bar{\xi}$ and $a$. \citet{Khan2008} showed that these parameters have little impact on the aggregate macro implications of the model in their preferred calibration; hence, micro data is needed. We keep all other parameters fixed at their true values for simplicity. \cref{sec:example_firm_prod} provides results for an alternative exercise where we instead estimate the parameters of the firms' idiosyncratic productivity process; the key messages are qualitatively similar to those presented below.

We adopt the annual calibration of \citet{Winberry2018}, which in turn follows \citet{Khan2008}, see \cref{sec:example_firm_calib}. However, we make an exception in setting the firm's idiosyncratic log productivity AR(1) parameter $\rho_\epsilon=0.53$, following footnote 5 in \citet{Khan2008}.\footnote{This avoids numerical issues that arise when solving the model for high degrees of persistence, as required in the estimation exercise in \cref{sec:example_firm_prod}.} We then adjust the log productivity innovation standard deviation $\sigma_\epsilon=0.0364$, so that the variance of the idiosyncratic log productivity process is unchanged from the baseline calibration in \citet{Khan2008} and \citet{Winberry2018}. The macro implications of our calibration are virtually identical to the baseline in \citet{Khan2008}, as those authors note.

We assume that the econometrician observes time series on aggregate output and investment, as well as repeated cross sections of micro data on firms' capital and labor inputs. We simulate macro data with sample size $T=50$, while micro cross sections of size $N=1000$ are observed at each point in time $t=1,\dots,50$. Unlike in \cref{sec:example_hh}, we do not add measurement error to the macro observables.

The prior on $(\bar{\xi},a)$ is chosen to be flat in the natural parameter space.

\subsection{Computation}
As in \cref{sec:example_hh}, we solve and simulate the model using the \citet{Winberry2018} Dynare solution method. We follow \citet{Winberry2018} and approximate the cross-sectional density of the firms' micro state variables (log capital and idiosyncratic productivity) with a multivariate normal distribution. Computation of the micro sampling density is simple, since -- conditional on macro states -- the micro observables (capital and labor) are log-linear transformations of these micro state variables. We use $J=500$ smoothing draws to compute the unbiased likelihood estimate. The MCMC routine is the same as in \cref{sec:example_hh}. The starting values are selected by a rough grid search on the simulated data. We generate 10,000 draws and discard the first 1,000 as burn-in. Likelihood evaluation using 20 parallel cores is several times faster than computing the model's steady state.

\subsection{Results}
\label{sec:example_firm_results}
\begin{figure}[t]
\centering
\textsc{Heterogeneous firm model: Posterior density, multiple simulations} \\[\baselineskip]
\includegraphics[width=\linewidth]{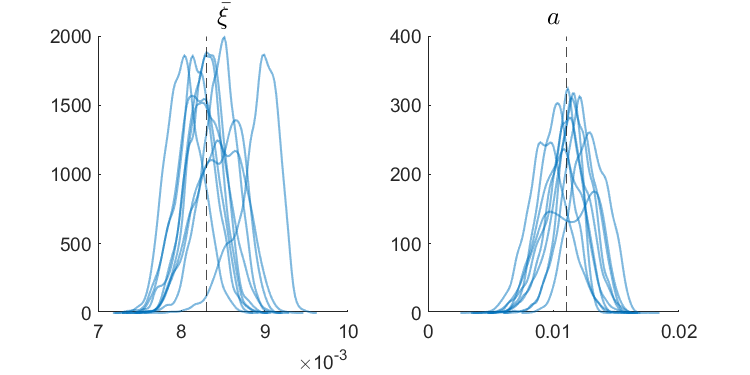}
\caption{Posterior densities across 10 simulated data sets. Vertical dashed lines indicate true parameter values. Posterior density estimates from the 9,000 retained MCMC draws using Matlab's {\tt ksdensity} function with default settings.}
\label{fig:example_firm_postdens}
\end{figure}

Despite the finding in \citet{Khan2008} that macro data is essentially uninformative about the firms' adjustment cost parameters, these are accurately estimated when the micro data is used also. \cref{fig:example_firm_postdens} shows the posterior densities of $\bar{\xi}$ and $a$ computed on 10 different simulated data sets. The posterior distribution of each parameter is systematically concentrated close to the true parameter values. We refrain from visually comparing these results with inference that relies only on macro data, since the macro likelihood is almost entirely flat as a function of $(\bar{\xi},a)$, consistent with \citet{Khan2008}.\footnote{On average across the 10 simulated data sets, the standard deviation (after burn-in) of the macro log likelihood $\log p(\mathbf{x} \mid \theta)$ across all Metropolis-Hastings proposals of the parameters is only 0.14, while it is 18.7 for the micro log likelihood $\log p(\mathbf{y} \mid \mathbf{x}, \theta)$.} Thus, micro data is essential to inference about these parameters. This finding is broadly consistent with \citet{Bachmann2014}, who show that the dynamics of the cross-sectional dispersion of firm investment are very informative about the nature of firm-level frictions.

\subsection{Correcting for imperfect sampling of micro data}
\label{sec:example_firm_censor}
One advantage of the likelihood approach adopted in this paper is that standard techniques can be applied to correct for sample selection or censoring in the micro data. This is highly relevant for applied work, since household or firm surveys are often subject to known data imperfections, even beyond measurement error.

Valid inference about structural parameters merely requires that the micro sampling density $p(y_{i,t} \mid z_t,\theta)$ introduced in \cref{sec:framework} accurately reflects the sampling mechanism, including the effects of selection or censoring. Hence, if it is known, say, that an observed variable such as household income is top-coded (i.e., censored) at the threshold $\bar{y}$, then the functional form of the density $p(y_{i,t} \mid z_t,\theta)$ should take into account that the observed data equals a transformation $y_{i,t} = \min\lbrace \tilde{y}_{i,t},\bar{y} \rbrace$ of the theoretical household income $\tilde{y}_{i,t}$ in the DSGE model. The likelihood functions of such limited dependent variable sampling models are well known and readily looked up, see for example \citet[chapters 17 and 19]{Wooldridge2010}.\footnote{If the nature of the data imperfection is only partially known, it may be possible to estimate the sampling mechanism from the data. For example, if the data is suspected to be subject to endogenous sample selection, one could specify a Heckman-type selection model and estimate the parameters of the selection model as part of the likelihood framework \citep[chapter 19]{Wooldridge2010}. It is outside the scope of this paper to consider nonparametric approaches or to analyze the consequences of misspecification of the sampling mechanism.} We provide one illustration below.

Other approaches to estimating heterogeneous agent models do not handle data imperfections as easily or efficiently. For example, inference based on cross-sectional moments of micro observables may require lengthy derivations to adjust the moment formulas for selection or censoring, especially for higher moments. Moreover, even in models where low-dimensional sufficient statistics exist for the underlying micro variables, cf.\ \cref{sec:theory_moment}, the \emph{imperfectly observed} micro data may not afford such sufficient statistics. In contrast, our likelihood-based approach is automatically efficient, and the adjustments needed to account for common types of data imperfections can be looked up in microeconometrics textbooks.

\paragraph{Illustration: Selection on outcomes.}
We illustrate the previous points by adding an endogenous selection mechanism to the sampled micro data in the heterogeneous firm model. Assume that instead of observing a representative sample of firms every period, we observe the draws for those firms whose employment in that period exceeds the 90th percentile of the steady-state cross-sectional distribution of employment. To make the effective micro sample size comparable to that in \cref{sec:example_firm_results}, we here set the per-period micro sample size \emph{before selection} equal to $N=10000$. That is, out of 10000 potential draws in a period, we only observe the capital and labor inputs of the approximately 1000 largest firms. Though stylized, this sampling mechanism is intended to mimic the real-world phenomenon that databases such as Compustat tend to only cover the largest active firms in the economy.

To adjust the likelihood for selection, we combine the model-implied cross-sectional distribution of the idiosyncratic state variables with the functional form of the selection mechanism. Let $g_t(\epsilon,k)$ be the cross-sectional distribution of idiosyncratic log productivity $\epsilon_{i,t}$ and log capital $k_{i,t}$ at time $t$, implied by the model (this density is approximated using an exponential family of densities, as in \citealp{Winberry2018}). In the model, log employment is given by $n_{i,t} = (\log\nu + \zeta_t - \log(w_t) + \epsilon_{i,t} + \alpha k_{i,t})/(1-\nu)$, where $w_t$ is the aggregate wage, $\zeta_t$ is log aggregate TFP, and $\nu$ and $\alpha$ are the output elasticities of labor and capital in the firm production function ($\nu+\alpha<1$). Since observations $y_{i,t} = (n_{i,t},k_{i,t})'$ are observed if and only if $n_{i,t} \geq \bar{n}$, the micro sampling density is given by the truncation formula\footnote{The integral in the denominator can be computed in closed form if the density $g_t(\epsilon,k)$ is multivariate Gaussian, which is the approximation we use in our numerical experiments, following \citet{Winberry2018}.}
\[p(n_{i,t},k_{i,t} \mid z_t,\theta) = \frac{(1-\nu) g_t\big((1-\nu)n_{i,t} - \alpha k_{i,t} - \log\nu - \zeta_t + \log(w_t), k_{i,t} \big)}{\int_{-\infty}^\infty \int_{-\infty}^\infty \mathbbm{1}\big(\log\nu + \zeta_t - \log(w_t) + \epsilon + \alpha k \geq (1-\nu)\bar{n}\big) g_t(\epsilon,k) \, d\epsilon \, dk}.\]
The selection threshold $\bar{n}$ is given by the true 90th percentile of the steady-state distribution of log employment. We assume this threshold is known to the econometrician for simplicity.\footnote{In principle, $\bar{n}$ could be treated as another parameter to be estimated from the available data.}

\begin{figure}[tp]
\centering
\textsc{Heterogeneous firm model: Posterior densities with selection} \\[\baselineskip]
\includegraphics[width=\linewidth]{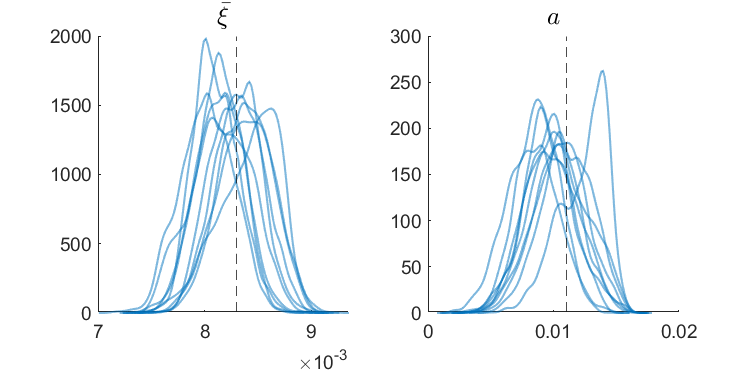}
\caption{Posterior densities across 10 simulated data sets subject to selection. Vertical dashed lines indicate true parameter values. Posterior density estimates from the 9,000 retained MCMC draws using Matlab's {\tt ksdensity} function with default settings.}
\label{fig:example_firm_postdens_trunc}
\end{figure}

\cref{fig:example_firm_postdens_trunc} shows the posterior distribution of the adjustment cost parameters $(\bar{\xi},a)$ in the model with selection, across 10 simulated data sets. All settings are the same as in \cref{sec:example_firm_results}, except for (i) the selection mechanism in the simulated micro data and the requisite adjustment to the functional form of the micro likelihood function, and (ii) the pre-selection micro sample size $N=10000$ (as discussed above). The posterior distributions of the parameters of interest remain centered close to the true parameter values, with no appreciable increase in posterior uncertainty relative to \cref{fig:example_firm_postdens}. This example demonstrates that data imperfections can be handled in a valid and efficient manner using standard likelihood techniques.

\section{Extension to panel data}
\label{sec:panel}
While our baseline procedure in \cref{sec:method} assumes the micro data to be given by repeated cross sections, we now consider settings where the micro data has a panel dimension -- that is, the same cross-sectional units are observed over two or more consecutive time periods. For tractability, we focus on panel data sets where the time dimension per unit is short (similar to \citealp{Papp2020}). One example is rotating panel survey data, where each unit is observed for a few consecutive time periods, after which it is replaced by a new, representatively sampled unit (as in the Bureau of Labor Statistics' Consumer Expenditure Survey). Panel data sets with a large time dimension, such as detailed administrative data sets, are computationally challenging and beyond the scope of this paper.

\subsection{Challenges and solutions}


The main challenge in handling panel data is the need to integrate out any unobserved individual-specific state variables (such as idiosyncratic productivity or asset holdings) that influence agents' dynamic decision rules. If the structural model directly implies a simple functional form for the one-step-ahead predictive density $p(y_{i,t} \mid y_{i,t-1},y_{i,t-2},\dots,\mathbf{z},\theta)$ of the \emph{observed} data $y_{i,t}$ for individual $i$, then evaluating the micro likelihood is trivial (as in reduced-form dynamic panel data models). Unfortunately, in most settings this predictive density is not available in closed form, and must instead be computed as the integral $\int p(y_{i,t} \mid s_{i,t},\mathbf{z},\theta)p(s_{i,t} \mid y_{i,t-1},y_{i,t-2},\dots,\mathbf{z},\theta)\,ds_{i,t}$ over the latent micro state variables $s_{i,t}$. Whereas the first density in the integrand may often be available in closed form, the second density will typically not be (outside simple linear models). Hence, evaluating the integral for each $i$ and $t$ appears to be computationally infeasible in many applications, especially if the number of time periods is moderately large. Nevertheless, as we now show, it is often possible to evaluate the micro likelihood when the time dimension per unit is small, by avoiding direct evaluation of the intractable predictive density.

Our proposal for exploiting panel data utilizes the model-implied relationship between (i) the latent micro state variables $s_{i,t_i}$ in the \emph{initial} period $t_i$ for individual $i$ and (ii) the micro observables $\lbrace y_{i,t} \rbrace_t$ in all observed time periods for that individual. This relation necessarily involves iterating on the dynamic micro decision rules of the agents in the economy. In the next subsection, we explain this approach by example, using the heterogeneous household model from \cref{sec:model_hh}. Since the main focus of this paper is repeated cross-sectional micro data, we leave a numerical exploration of the benefits of panel data for future work.

\subsection{Example: Heterogeneous household model} \label{sec:panel_example}
Unlike in \cref{sec:model_hh}, we here assume that we observe two consecutive periods of $y_{i,t}=(\epsilon_{i,t},\iota_{i,t})$ for each household $i$, i.e., household employment and income. To implement the likelihood estimate in \cref{sec:method}, we must evaluate the conditional micro density
\[p(y_{i,t},y_{i,t-1} \mid \mathbf{z},\theta) = \underbrace{p(\epsilon_{i,t},\epsilon_{i,t-1} \mid \mathbf{z},\theta)}_{=p(\epsilon_{i,t-1},\theta)p(\epsilon_{i,t} \mid \epsilon_{i,t-1},\theta)} p(\iota_{i,t},\iota_{i,t-1} \mid \epsilon_{i,t},\epsilon_{i,t-1},\mathbf{z},\theta).\]
Employment $\epsilon_{i,t}$ evolves as a simple exogenous two-state Markov process, so the challenge is to evaluate the last density on the right-hand side above.

Note that, by definition of household income,
\[\iota_{i,t-1} = \lambda_i \big( w_{t-1}[(1-\tau)\epsilon_{i,t-1} + b(1-\epsilon_{i,t-1})] + (1+r_{t-1})a_{i,t-2} \big),\]
and
\[\iota_{i,t} = \lambda_i \big( w_t[(1-\tau)\epsilon_{i,t} + b(1-\epsilon_{i,t})] + (1+r_{t})a_{t-1}'(a_{i,t-2},\epsilon_{i,t-1})\big),\]
where $a_{t-1}'(a,\epsilon)$ is the model-implied micro policy function at period $t-1$ for next-period normalized assets given current normalized assets $a$ and current employment $\epsilon$, and given the aggregate state $z_{t-1}$.

Conditional on $(\epsilon_{i,t},\epsilon_{i,t-1})$ and the macro states $\mathbf{z}$ in all periods, the observation $(\iota_{i,t},\iota_{i,t-1})$ is therefore a known transformation of the initial micro state vector $s_{i,t-1} = (\lambda_i,a_{i,t-2})$.\footnote{Strictly speaking, employment $\epsilon_{i,t-1}$ is also a micro state variable. However, since it follows an exogenous Markov process, our derivations above condition on it, and we can therefore disregard it in $s_{i,t-1}$.} We can then derive $p(\iota_{i,t},\iota_{i,t-1} \mid \epsilon_{i,t},\epsilon_{i,t-1},\mathbf{z},\theta)$ by applying the change-of-variables formula to the density $p(\lambda_i,a_{i,t-2} \mid \epsilon_{i,t},\epsilon_{i,t-1},\mathbf{z},\theta) = f(\lambda_i \mid \mu_\lambda)g_{\epsilon_{i,t-1}}(a_{i,t-2} \mid \psi_{t-1})$, where $\psi_{t-1}$ denotes the parameters governing the cross-sectional distribution and is part of the aggregate state $z_{t-1}$, and the density $g_{\epsilon_{i,t-1}}(a_{i,t-2} \mid \psi_{t-1})$ is directly available from the output of the model solution procedure, as discussed in \cref{sec:model_hh}.\footnote{Note that assets are predetermined, so the subscript for the distribution parameters is $t-1$.} Computing the Jacobian term in the change-of-variables formula requires us to evaluate the derivative $\partial a_{t-1}'(a,\epsilon)/\partial a$, for example by applying finite differences to the function $a_{t-1}'(a,\epsilon)$ that is outputted by the numerical model solution method.\footnote{If the solution method uses an approximate, discrete grid for this function, one possibility is to compute the derivative of a smooth interpolation of the discretized rule.}

\subsection{Summary and discussion}
In general terms, our proposal is to express the consecutive micro observations in terms of the latent micro state variables $s_{i,t_i}$ in the initial observed period $t_i$ for individual $i$. We can then ``invert'' this relation and evaluate the micro likelihood using the model-implied cross-sectional density of $s_{i,t_i}$ (which we also exploited previously in the repeated cross section setup). This strategy is highly context-specific as it exploits the structure of the observables and iterates on the model-implied dynamic decision rules of the agents. Though we have only illustrated the strategy for the case of two-period panel data, the idea could in principle be applied to longer panels by further iterating on the decision rules; however, this could become cumbersome when the time dimension is moderately large. As a side note, it is straightforward to allow for measurement error in the micro observables by simply adding independent noise to the density of the noise-less observables using the convolution formula.\footnote{Unlike the repeated cross section case, in the case of panel data, unobserved individual heterogeneity and micro measurement error are not observationally equivalent.}

In some models the dynamic relationship between micro states and micro observables may be sufficiently convoluted to render the above approach impractical. For such cases, we propose an alternative approach in \cref{sec:panel_general} based on artificially adding lagged state variables to the micro state vector in the numerical model solution procedure. Though this alternative approach is conceptually simple to implement, the increase in the dimension of the effective micro state vector may require more time to be spent on computing an accurate numerical solution of the model. We therefore recommend that researchers first attempt the baseline approach illustrated in the previous subsection, which does not require any modification of the numerical model solution method.

\section{Conclusion}
\label{sec:concl}
The literature on heterogeneous agent models has hitherto relied on estimation approaches that exploit \emph{ad hoc} choices of micro moments and macro time series for estimation. This contrasts with the well-developed framework for full-information likelihood inference in \emph{representative agent} models \citep{Herbst2016}. We develop a method to exploit the full information content in macro and micro data when estimating heterogeneous agent models. As we demonstrate through economic examples, the \emph{joint} information content available in micro and macro data is often much larger than in either of the two separate data sets. Our inference procedure can loosely be interpreted as a two-step method: First we estimate the underlying macro states from macro data, and then we evaluate the likelihood by plugging into the cross-sectional sampling densities given the estimated states. However, our method delivers finite-sample valid and fully efficient Bayesian inference that takes into account all sources of uncertainty about parameters and states. The computation time of our procedure scales well with the size of the data set, as the method lends itself to parallel computing. Unlike estimation approaches based on tracking a small number of cross-sectional moments over time, our full-information method is automatically efficient and can easily accommodate unobserved individual heterogeneity, micro measurement error, as well as data imperfections such as censoring or selection.

For clarity, we have limited ourselves to numerical illustrations with small-scale models in this paper, leaving full-scale empirical applications to future work. Our approach is computationally most attractive when the model is solved using some version of the \citet{Reiter2009} linearization approach, since this yields simple formulas for evaluating the macro likelihood and drawing from the smoothing distribution of the latent macro states, cf.\ \cref{sec:method}. To estimate large-scale quantitative models it would be necessary to apply now-standard dimension reduction techniques or other computational shortcuts to the linearized representation of the macro dynamics \citep{Ahn2017,Auclert2020seq}, and we leave this to future research. Nevertheless, we emphasize that our method is in principle generally applicable, as long as there exists some way to evaluate the macro likelihood, draw from the smoothing distribution of the macro states, and evaluate the micro sampling density given the macro states.

Our research suggests several additional avenues for future research. First, it would be useful to go beyond our extension to short panel data sets in \cref{sec:panel} and develop methods that are computationally feasible when the time dimension of the panel is large.  Second, since our method works for a wide range of generic MCMC posterior sampling procedures, it would be interesting to investigate the scope for improving on the simple Random Walk Metropolis-Hastings algorithm that we use for conceptual clarity in our examples. Third, the goal of this paper has been to fully exploit all aspects of the assumed heterogeneous agent model when doing statistical inference; we therefore ignore the consequences of misspecification. Since model misspecification potentially affects the entire macro equilibrium and thus cannot be addressed using off-the-shelf tools from the microeconometrics literature, we leave the development of robust inference approaches to future work.

\clearpage

\appendix

\section{Appendix}
\label{sec:appendix}

\subsection{Proofs}
\label{sec:moment_proof}

\subsubsection{Proof of \texorpdfstring{\cref{thm:suff_stat}}{Theorem \ref{thm:suff_stat}}}

Let $\hat m_t \equiv (\hat m_{1,t},\dots,\hat m_{Q,t})'$ denote the set of sufficient statistics in period \(t\). According to the Fisher-Neyman factorization theorem, there exists a function \(h(\cdot )\) such that the likelihood of the micro data in period $t$, conditional on $z_t$, can be factorized as
\begin{align}\prod_{i=1}^{N_t} p(y_{i,t} \mid z_t,\theta)
&=h(y_t)p(\hat m_t \mid N_t,z_t,\theta).\label{lik-t}
\end{align}
Let $\mathbf {h}(\mathbf{y})=\prod_{t\in\mathcal T}h(y_t)$, $\mathbf N = \{N_t\}_{t\in\mathcal T}$, and $\mathbf{\hat m} =\{\hat m_t\}_{t\in\mathcal T}$, where $\mathcal T$ is the subset of time points with observed micro data. Then the micro likelihood, conditional on the observed macro data, can be decomposed as
\begin{align}
p(\mathbf{y} \mid \mathbf{x},\theta) &= \int p(\mathbf{y} \mid \mathbf{z}, \theta) p(\mathbf{z} \mid \mathbf{x},\theta)\, d\mathbf{z}\nonumber\\
&=\mathbf {h}(\mathbf{y}) \int p(\mathbf{\hat m} \mid \mathbf N, \mathbf{z}, \theta) p(\mathbf{z} \mid \mathbf{x},\theta)\, d\mathbf{z}\label{eqn:micro-lik}\\
&=\mathbf {h}(\mathbf{y}) p(\mathbf{\hat m} \mid \mathbf N, \mathbf{x},\theta). \label{eqn:micro-lik2}
\end{align}
The expression \eqref{eqn:micro-lik2} implies that \(\mathbf{\hat m}\) is a set of sufficient statistics for $\theta$, based again on the Fisher-Neyman factorization theorem. \qed

\subsubsection{Proof of \texorpdfstring{\cref{cor:exp_poly}}{Corollary \ref{cor:exp_poly}}}

Let $m_t$ be a vector of population counterparts of the cross-sectional sufficient statistics of the micro states $s_{i,t}$. We may view $m_t$ as part of the macro state vector $z_t$. According to the exponential polynomial setup,
\begin{align*}p(s_{i,t} \mid z_t,\theta)&=p(s_{i,t} \mid m_t)\\
&=\exp\left[\tilde \varphi_0(m_t)+\sum_{\ell=1}^{Q} \tilde \varphi_\ell(m_t)\tilde {\mathfrak m}_{\ell}(s_{i,t})\right].
\end{align*}
$\tilde {\mathfrak m}_{\ell}(s_{i,t})$ takes the form $s_{i,t,1}^{p_1} s_{i,t,2}^{p_2}\cdots s_{i,t,d_s}^{p_{d_s}}$ with $p_k$ being positive integers and $1\le\sum_{k=1}^{d_s}p_k\le q$, where $q$ is the order of the exponential polynomial. The potential number of sufficient statistics $Q$ equals $\binom{q+d_s}{q}-1$, i.e., the number of complete homogeneous symmetric polynomials.

Making the change of variables in \eqref{eqn:transform}, we have
\begin{align*}
p(y_{i,t} \mid z_t,\theta)&=\exp\left[\tilde \varphi_0(m_t)+\sum_{\ell=1}^{Q} \tilde \varphi_\ell(m_t)\tilde {\mathfrak m}_{\ell}\big(B_1(z_t,\theta)\Upsilon(y_{i,t})+B_0(z_t,\theta)\big)\right] \\
&\qquad \times \left|\det\left(B_1(z_t,\theta)\frac{\partial \Upsilon(y_{i,t})}{\partial y_{i,t}}\right)\right|\\
&\equiv\exp\left[\varphi_0(z_t,\theta)+{\mathfrak m}_{0}(y_{i,t})+\sum_{\ell=1}^{Q} \varphi_\ell(z_t,\theta){\mathfrak m}_{\ell}(y_{i,t})\right].
\end{align*}
Given assumptions 2.a and 2.b, the potential number of sufficient statistics $Q$ remains the same. Now the sufficient statistics can be expressed as ${\mathfrak m}_{\ell}(y_{i,t})\equiv\tilde {\mathfrak m}_{\ell}(\Upsilon(y_{i,t}))$ and the corresponding $\varphi_\ell(z_t,\theta)$ can be obtained by rearranging terms and collecting coefficients on ${\mathfrak m}_{\ell}(y_{i,t})$. For the determinant of the Jacobian, condition 2 implies that both $B_1(z_t,\theta)$ and $\frac{\partial \Upsilon(y_{i,t})}{\partial y_{i,t}}$ are non-singular square matrices, so \(\det\left(B_1(z_t,\theta)\frac{\partial \Upsilon(y_{i,t})}{\partial y_{i,t}}\right)=\det\left(B_1(z_t,\theta)\right)\det\left(\frac{\partial \Upsilon(y_{i,t})}{\partial y_{i,t}}\right)\). Hence,  both $\log\left|\det\left(B_1(z_t,\theta)\right)\right| $ and $\log\left|\det\left(\frac{\partial \Upsilon(y_{i,t})}{\partial y_{i,t}}\right)\right| $ are finite and can be absorbed into $\varphi_0(z_t,\theta)$ and ${\mathfrak m}_{0}(y_{i,t}) $, respectively. 

Thus, the micro likelihood fits into the general form in \cref{thm:suff_stat}, and the sufficient statistics are given by
\[\pushQED{\qed}
\hat m_{\ell,t} = \frac 1{N_t} \sum_{i=1}^{N_t} {\mathfrak m}_{\ell}(y_{i,t})=\frac 1{N_t} \sum_{i=1}^{N_t}\tilde {\mathfrak m}_{\ell}(\Upsilon(y_{i,t})), \quad \ell=1,\dots,Q. \qedhere\]

\subsection{Non-existence of sufficient statistics: Details}
\label{sec:suff_stat_general}
Can we generalize beyond the sufficient conditions in \cref{cor:exp_poly}? The key is that in \eqref{eqn:exp}, the terms inside the exponential should be additive and each term should take the form \(\varphi_\ell(z_t,\theta){\mathfrak m}_\ell(y_{i,t})\), which ensures that the cross-sectional moments can be calculated using micro data as in equation \eqref{eqn:moments} and the multiplicative term $\mathbf h(\mathbf{y}) $ can be taken out of the integral in equation \eqref{eqn:micro-lik}.
Building on the analysis of \cref{sec:theory_moment}, here are more details regarding cases where there are no sufficient statistics in general. \begin{enumerate}[i)]
\item $s_{i,t}=B_1(z_t,\theta)\Upsilon(y_{i,t},z_{t})+B_0(z_t,\theta)$, i.e., $y_{i,t}$ and $z_{t} $ are neither additively nor multiplicatively separable.
\item The model features unobserved individual heterogeneity and/or micro measurement error.
Since these two cases are observationally equivalent in a repeated cross section framework, we focus on the the former. Letting $\lambda_i$ denote the unobserved individual heterogeneity, we can extend \eqref{eqn:transform} to $s_{i,t}=\tilde \Upsilon(y_{i,t},\lambda_i,z_{t},\theta) \equiv B_1(\lambda_i,z_t,\theta)\Upsilon(y_{i,t},\lambda_{i})+B_0(\lambda_i,z_t,\theta)$, which is the most general setup allowing $\lambda_i$ to affect all terms in the expression.
If $\lambda_i$ is independent of $s_{i,t}$ conditional on \((z_{t},\theta)\), we have $p(s_{i,t},\lambda_i \mid z_t,\theta)=p(s_{i,t} \mid m_t)p(\lambda_i\mid\theta)$ (recall the notation in the proof of \cref{cor:exp_poly}). Accordingly,
\[p(y_{i,t} \mid z_t,\theta)=\int p(\tilde \Upsilon(y_{i,t},\lambda_i,z_{t},\theta) \mid m_t)\left|\det\left(B_1(\lambda_i,z_t,\theta)\frac{\partial \Upsilon(y_{i,t},\lambda_i)}{\partial y_{i,t}}\right)\right|p(\lambda_i\mid\theta)\, d\lambda_i.\]
If $\lambda_i$ appears in $B_1$, $B_0$, or $\Upsilon$, then $p(y_{i,t} \mid z_t,\theta)$ may not belong to the exponential family after integrating out $\lambda_i$.
That said, we can construct special cases where sufficient statistics do exist. For example, if $s_{i,t}=B_1(z_t,\theta)\Upsilon(y_{i,t})+B_0(z_t,\theta)+B_2(z_t,\theta)\lambda_i$ and both $p(s_{i,t} \mid m_t)$ and $p(\lambda_i\mid\theta)$ follow Gaussian distributions.
\item $d_s>d_y$: For example, suppose $s_{i,t}$ is two-dimensional whereas $y_{i,t}$ is one-dimensional, say $y_{i,t}=s_{1,i,t}$, $y_{i,t}=s_{1,i,t}+s_{2,i,t}$, or $y_{i,t}=s_{1,i,t}s_{2,i,t}$. We can first expand the $y_{i,t}$ in \eqref{eqn:transform} to $\tilde y_{i,t} = (y_{i,t},s_{2,i,t})'$ and then integrate out $s_{2,i,t}$. However, after the integration, the resulting micro likelihood as a function of $y_{i,t}$ may not take the exponential family form anymore.
\end{enumerate}

\subsection{Sampling distribution of cross-sectional moments: Example}
\label{sec:nonlinear-Gaussian}
As alluded to in \cref{sec:theory_moment}, here is a simple example demonstrating that $p(\hat m_t\mid N_t, z_t,\theta)$ is not linear Gaussian in finite samples, and therefore neither is $ p(\mathbf{\hat m} \mid \mathbf N, \mathbf{x},\theta)$. Suppose $y_{i,t}=s_{i,t}$ is a scalar and $p(s_{i,t} \mid m_t)$ is Gaussian, i.e., a second-order exponential polynomial. Let $\hat m_{1,t}=\frac 1 {N_t}\sum_{i=1}^{N_t} s_{i,t}$ and $\hat m_{2,t}=\frac 1 {N_t}\sum_{i=1}^{N_t} (s_{i,t}-\hat m_{1,t})^2$ with $m_{1,t}$ and $m_{2,t}$ being their population counterparts. Then standard calculations yield
\begin{align*}
p(\hat m_t\mid N_t, z_t,\theta)=p(\hat m_t \mid m_t )=\phi\left(\hat m_{1,t};\,m_{1,t},\frac{m_{2,t}}{N_t}\right)p_{\chi^2}\left(\frac {N_t \hat m_{2,t}} {m_{2,t}};\,N_t-1 \right),
\end{align*}
where $\phi(x;\mu,\sigma^2)$ represents the probability distribution function (pdf) of a Gaussian distribution with mean $\mu$ and variance $\sigma^2$, and $p_{\chi^2}(x;\nu)$ is the pdf of a chi-squared distribution with $\nu$ degrees of freedom. We can see that the latter is not linear Gaussian. Moreover, when $p(s_{i,t} \mid m_t)$ follows a higher order exponential polynomial, the characterization of $p(\hat m_t\mid N_t, z_t,\theta)$ would be even more complicated without a closed-form expression.

\subsection{Heterogeneous household model: Likelihood comparison}
\label{sec:example_hh_lik_multi}
\begin{figure}[p]
\centering
\textsc{Het.\ household model: Likelihood comparison, multiple simulations} \\[0.5\baselineskip]
\begin{tabular}{cc}
&\hspace{.15in}$\beta$\hspace{1.35in}$\sigma_e$\hspace{1.35in}$\mu_\lambda$\\
\rotatebox{90}{\hspace{.5in} Full Info} &
\adjustbox{trim={.08\width} {0\height} {.05\width} {0\height},clip}%
  {\includegraphics[width=5.4in]{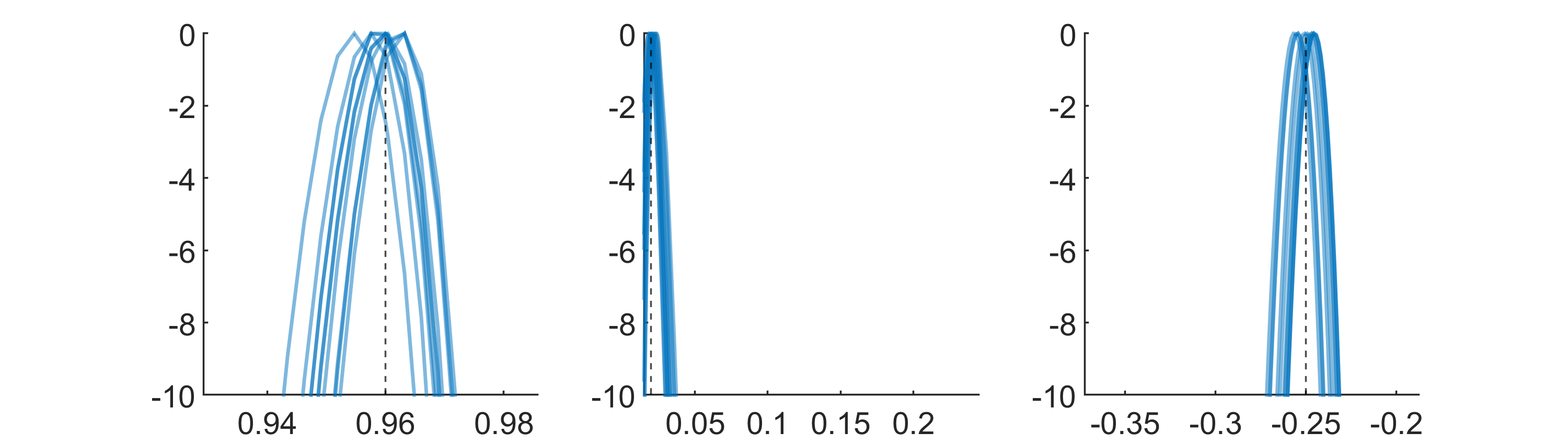}} \\
  \rotatebox{90}{\hspace{.4in} Macro Only} &
\adjustbox{trim={.08\width} {0\height} {.05\width} {0\height},clip}%
  {\includegraphics[width=5.4in]{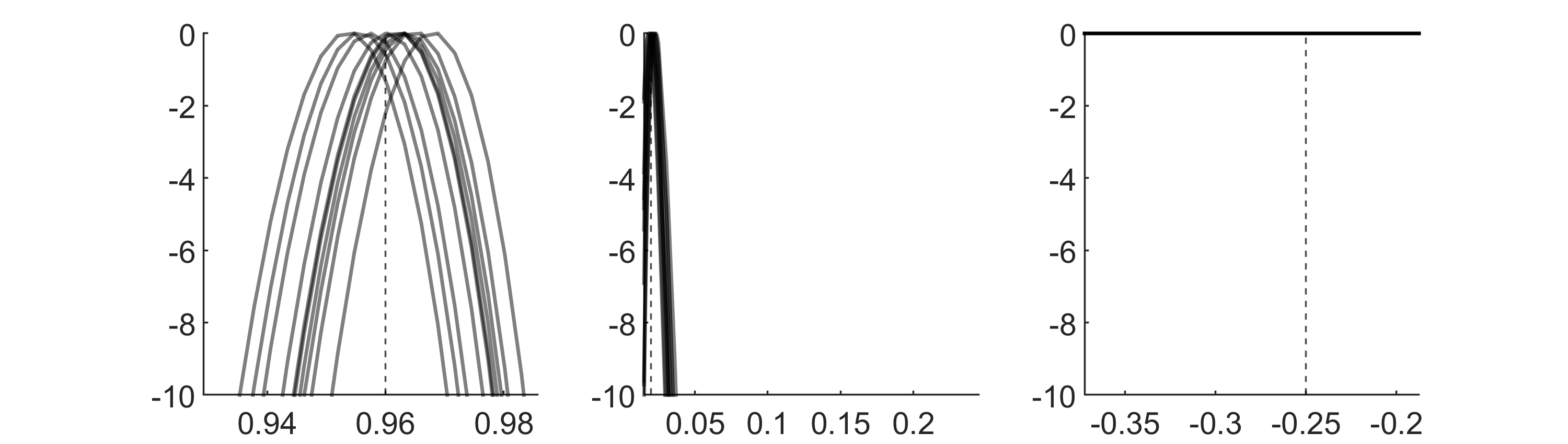}} \\
  \rotatebox{90}{\hspace{.3in} 3rd Moment} &
\adjustbox{trim={.08\width} {0\height} {.05\width} {0\height},clip}%
  {\includegraphics[width=5.4in]{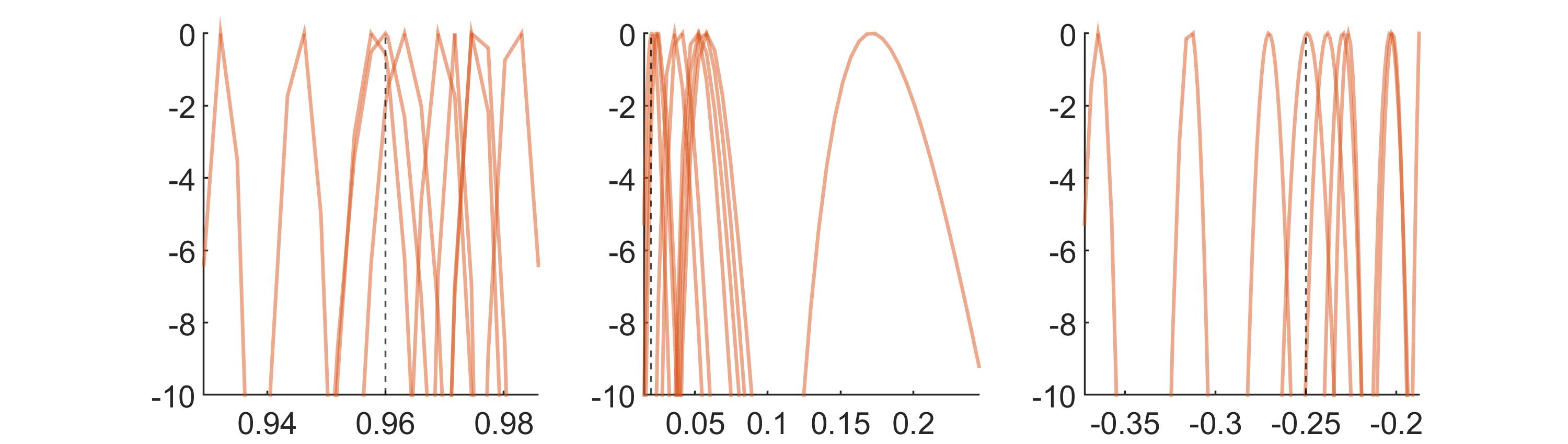}} \\
  \rotatebox{90}{\hspace{.3in} 2nd Moment} &
\adjustbox{trim={.08\width} {0\height} {.05\width} {0\height},clip}%
  {\includegraphics[width=5.4in]{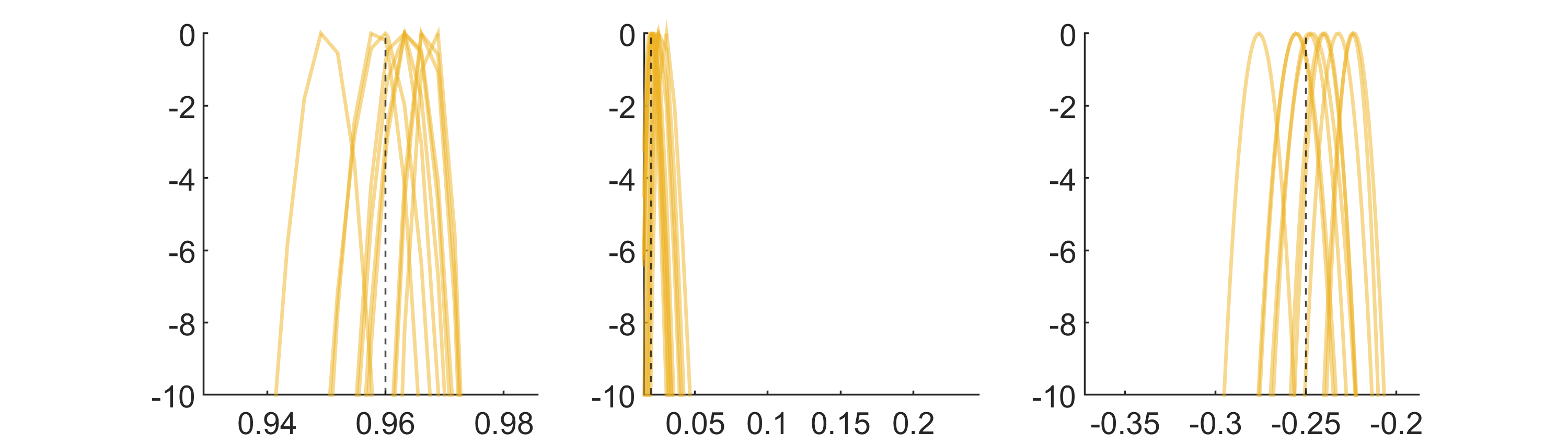}} \\
  \rotatebox{90}{\hspace{.3in} 1st Moment} &
\adjustbox{trim={.08\width} {0\height} {.05\width} {0\height},clip}%
  {\includegraphics[width=5.4in]{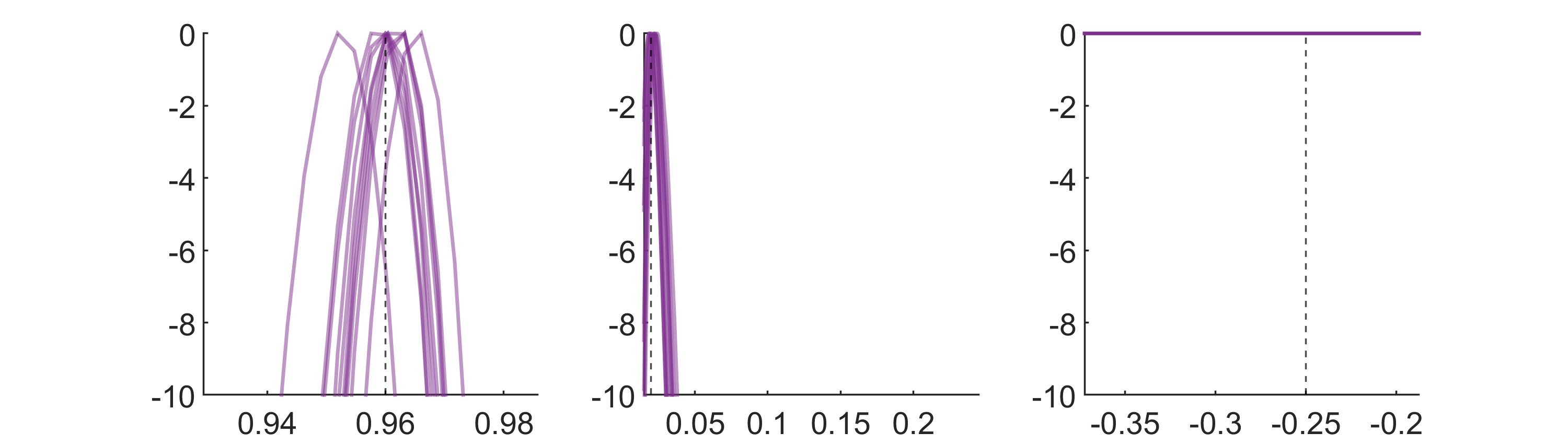}} 
\end{tabular}
\caption{Comparison of log likelihood functions across 10 different simulated data sets. See the description in \cref{sec:example_hh_lik_multi}.}
\label{fig:example_hh_lik_byest}
\end{figure}

Complementing the results for a single simulated data set in \cref{sec:example_hh_lik}, \cref{fig:example_hh_lik_byest} compares log likelihoods for the different inference methods across 10 different simulated data sets. Here different inference methods are exhibited in different rows. Similar to \cref{fig:example_hh_lik_byrep}, each column depicts univariate deviations of a single parameter while keeping all other parameters at their true values. There are 10 likelihood curves in each panel, corresponding to the 10 simulated data sets. The maximum of each likelihood curve is normalized to be zero. Vertical dashed lines indicate true parameter values. The ``1st Moment'' and ``Macro Only'' curves are flat on the right panels of the second and the last rows, since $\mu_\lambda$ is not identified from this data alone. We conclude from the figure that the full-information likelihood is systematically well-centered and tightly concentrated around the true parameter values, whereas the various moment-based likelihoods are poorly centered, exhibit less curvature, and/or shift around substantially across simulations.

\clearpage

\phantomsection
\addcontentsline{toc}{section}{References}
\bibliography{het_agents_ref}

\end{document}


\title{Online Appendix:\texorpdfstring{\\}{}Full-Information Estimation of Heterogeneous Agent Models Using Macro and Micro Data}
\author{Laura Liu \and Mikkel Plagborg-M{\o}ller}
\date{\today}
\maketitle

\begin{appendices}
\crefalias{section}{sappsec}
\crefalias{subsection}{sappsubsec}
\crefalias{subsubsection}{sappsubsubsec}
\setcounter{section}{1}

\numberwithin{equation}{section}
\numberwithin{figure}{section}
\numberwithin{table}{section}

\section{Variance-covariance matrix for moment-based methods}
\label{sec:moment_vcv}
Here we describe how we estimate the variance-covariance matrix of the cross-sectional moments when implementing the moment-based inference approaches in \cref{sec:example_hh_lik}. We take the ``3rd Moment'' inference approach as an example. The measurement error variance-covariance matrices of other moment-based approaches can be derived in a similar fashion.

Let $\hat m_{\epsilon j,t}$, for $\epsilon=0,1$ and $j=1,2,3$, be the cross-sectional sample moments of household after-tax income in period $t$, and $m_{\epsilon j,t}$ be the corresponding population moments, where $\epsilon$ indicates the employment status of the group and $j$ represents the order of the moment, such as the sample mean, variance, and third central moment. For instance, \begin{align*}
\hat m_{11,t}&=\frac{\sum_{i=1}^{N}\iota_{i,t}\epsilon_{i,t}}{\sum_{i=1}^{N}\epsilon_{i,t}},\quad m_{11,t}=\mathbb E[\iota_{i,t}\mid \epsilon_{i,t}=1,\;z_t],\\
\hat m_{1j,t}&=\frac{\sum_{i=1}^{N}(\iota_{i,t}-\hat m_{11,t})^j\epsilon_{i,t}}{\sum_{i=1}^{N}\epsilon_{i,t}},\quad m_{1j,t}=\mathbb E[(\iota_{i,t}-m_{11,t})^j\mid \epsilon_{i,t}=1,\;z_t],\quad \text{for }j > 1.
\end{align*}
Define $\hat m_t \equiv \left(
\hat m_{01,t}, \hat m_{02,t}, \hat m_{03,t}, \hat m_{11,t}, \hat m_{12,t}, \hat m_{13,t} \right)'.$ To construct the measurement error variance-covariance matrix $\mathbb V\left[\hat m_t\mid z_t\right]$, we need to compute the variances and covariances across $\hat m_{\epsilon j,t}$'s. It is easy to see that $\hat m_{0j,t}$ and $\hat m_{1k,t}$ are asymptotically independent as $N\to\infty$ for any moment orders $j,k$, so we can focus on deriving the diagonal blocks where the moments share the same employment status $\epsilon$.

Let us first consider the variance of $\hat m_{11,t}$. As $(\iota_{i,t},\epsilon_{i,t})$ is cross-sectionally i.i.d.\ given $z_t$, we resort to the Central Limit Theorem and Slutsky's theorem and obtain\footnote{In the denominator, $\frac{1} {N}\sum_{i=1}^{N}\epsilon_{i,t}\overset{p}{\to} \mathbb{E}[\epsilon_{i,t}]$ as $N\to \infty$. Recall that $\mathbb{E}[\epsilon_{i,t}]=L$ for all $t$.}
\begin{equation}\mathbb V[\hat m_{11,t}\mid z_t] = \frac{\mathbb V[\iota_{i,t}\mid \epsilon_{i,t}=1,\;z_t]}{N L}+o_p\left(N ^{-1}\right).\label{eqn:v11}\end{equation}
The sample analog of $\mathbb V[\iota_{i,t}\mid \epsilon_{i,t}=1,\;z_t]$ is $\hat m_{12,t}$, the sample variance of the employed group in period $t$. As explained in the main text, we assume that the variance-covariance matrix of the moments is constant across time and estimate it using full-sample sample moments (i.e., averaging across time). Thus, the numerator in \eqref{eqn:v11} is approximated by $\hat m_{12} \equiv \frac 1 {|\mathcal T|}\sum_{t\in\mathcal T}\hat m_{12,t}$, where $\mathcal T$ is the subset of time points we observe the micro data, and $|\mathcal T|$ gives the number of elements in set $\mathcal T$.\footnote{In our numerical experiment, the micro sample size $N_t=N$ is constant over time. If this were not the case, $\hat{m}_{12}$ should be constructed using sample size weights.} Similarly, the denominator in \eqref{eqn:v11} can be approximated by $\hat N_1 \equiv \frac 1 {|\mathcal T|}\sum_{t\in\mathcal T}\sum_{i=1}^{N}\epsilon_{i,t}$.

For other elements in the ``employed'' block, we have
\begin{align*}
&\mathbb V[\hat m_{12,t}\mid z_t] = \frac{\mathbb V[\left(\iota_{i,t}- m_{11}\right)^2\mid \epsilon_{i,t}=1,\;z_t]}{N L}+o_p\left(N ^{-1}\right)\approx \frac{\hat m_{14}-\hat m_{12}^2}{\hat N_1},\\
&\mathbb V[\hat m_{13,t}\mid z_t] = \frac{\mathbb V[\left(\iota_{i,t}- m_{11}\right)^3\mid \epsilon_{i,t}=1,\;z_t]}{N L}+o_p\left(N ^{-1}\right)\approx \frac{\hat m_{16}-6\hat m_{14}\hat m_{12}-\hat m_{13}^2+9\hat m_{12}^3}{\hat N_1},\\
&\mathbb{C}\text{ov}[\hat m_{11,t},\hat m_{12,t}\mid z_t] = \frac{\text{cov}[\iota_{i,t}- m_{11},\;\left(\iota_{i,t}- m_{11}\right)^2\mid \epsilon_{i,t}=1,\;z_t]}{N L}+o_p\left(N ^{-1}\right)\approx \frac{\hat m_{13}}{\hat N_1},\\
&\mathbb{C}\text{ov}[\hat m_{11,t},\hat m_{13,t}\mid z_t] = \frac{\text{cov}[\iota_{i,t}- m_{11},\;\left(\iota_{i,t}- m_{11}\right)^3\mid \epsilon_{i,t}=1,\;z_t]}{N L}+o_p\left(N ^{-1}\right)\approx \frac{\hat m_{14}-3\hat m_{12}^2}{\hat N_1},\\
&\mathbb{C}\text{ov}[\hat m_{12,t},\hat m_{13,t}\mid z_t] = \frac{\text{cov}[\left(\iota_{i,t}- m_{11}\right)^2,\;\left(\iota_{i,t}- m_{11}\right)^3\mid \epsilon_{i,t}=1,\;z_t]}{N L}+o_p\left(N ^{-1}\right) \\
&\qquad\qquad\qquad\qquad\;\; \approx \frac{\hat m_{15}-4\hat m_{13}\hat m_{12}}{\hat N_1}.\end{align*}
In each equation, the first equality is given
by a similar cross-sectionally i.i.d.\ argument. The second approximation rewrites the variance/covariance in the numerator in terms of population moments \citep[as in][but omitting inconsequential degree-of-freedom adjustments]{fisher1930moments}, and then substitutes these population moments with their sample analogs averaged over time. Note that the last terms in the equations above call for even higher-order sample moments. Specifically, to approximate the variance/covariance involving the $m$-th and $n$-th order sample moments, we need sample moments up to the $(m+n)$-th order.

For the ``unemployed'' block, we can replace $\hat m_{1j,t},\;\epsilon_{i,t}=1,\;L,\;\hat m_{1j},\text{ and }\hat N_1$ with $\hat m_{0j,t},\;\epsilon_{i,t}=0,\;1-L,\;\hat m_{0j},\text{ and }\hat N_0 \equiv \frac 1 {|\mathcal T|}\sum_{t\in\mathcal T}\sum_{i=1}^{N}(1-\epsilon_{i,t}) $, respectively.

Combining all steps above, we can approximate the measurement error variance-covari\-ance matrix using sample moments of micro data:
\begin{align*}\mathbb V\left[\hat m_t\mid z_t\right]&\approx\begin{pmatrix}V_{00} & 0_{3\times3} \\
0_{3\times3} & V_{11} \\
\end{pmatrix},\\
V_{00}&\equiv \frac 1{\hat N_0}\left(\begin{array}{ccc}
\hat m_{02} & \hat m_{03} & \hat m_{04}-3\hat m_{02}^2\\
\hat m_{03} & \hat m_{04}-\hat m_{02}^2 & \hat m_{05}-4\hat m_{03}\hat m_{02} \\
\hat m_{04}-3\hat m_{02}^2 & \hat m_{05}-4\hat m_{03}\hat m_{02} & \substack{\hat m_{06}-6\hat m_{04}\hat m_{02}\\-\hat m_{03}^2+9\hat m_{02}^3} \\
\end{array}\right),\\
V_{11}&\equiv \frac 1{\hat N_1}\left(\begin{array}{ccc}
\hat m_{12} & \hat m_{13} & \hat m_{14}-3\hat m_{12}^2\\
\hat m_{13} & \hat m_{14}-\hat m_{12}^2 & \hat m_{15}-4\hat m_{13}\hat m_{12} \\
\hat m_{14}-3\hat m_{12}^2 & \hat m_{15}-4\hat m_{13}\hat m_{12} & \substack{\hat m_{16}-6\hat m_{14}\hat m_{12}\\-\hat m_{13}^2+9\hat m_{12}^3} \\
\end{array}\right).\end{align*}

\clearpage

\section{Heterogeneous household model}
\subsection{Calibration}
\label{sec:example_hh_calib}
\cref{tab:example_hh_calib} shows the parameter calibration used to simulate the data. Here $\pi(0 \to 1)$, for example, denotes the idiosyncratic Markov transition probability $P(\epsilon_{i,t+1}=1 \mid \epsilon_{i,t}=0)$.

\begin{table}[tp]
\centering
\textsc{Heterogeneous household model: Parameter calibration} \\[\baselineskip]
\renewcommand{\arraystretch}{1.1}
\begin{tabular}{lll|lll}
\hline
$\beta$  &  Discount factor & 0.96 & $\pi(0\rightarrow 1)$ & U to E trans. & 0.5 \\
$\alpha$ & Capital share & 0.36 &  $\pi(1\rightarrow 0)$ & E to U trans. & 0.038 \\
$\delta$ & Capital depreciation & 0.10 &  $\rho_\zeta$ & Agg.\ TFP AR(1) & 0.859 \\
$b$ & UI replacement rate & 0.15  &  $\sigma_\zeta$ & Agg.\ TFP AR(1) & 0.014 \\
$\mu_\lambda$ & Idiosyncratic distr. & -0.25 &  $\sigma_e$ & Meas.\ err.\ in output & 0.02\\
\hline             
\end{tabular}
\caption{Parameter calibration in the heterogeneous household model.} \label{tab:example_hh_calib}
\end{table}

\subsection{Additional simulation results}
\label{sec:example_hh_results}
\begin{figure}[tp]
\centering
\textsc{Het.\ household model: Consumption policy function, unemployed} \\[\baselineskip]
\includegraphics[width=\linewidth,clip=true,trim=2em 0 2em 0]{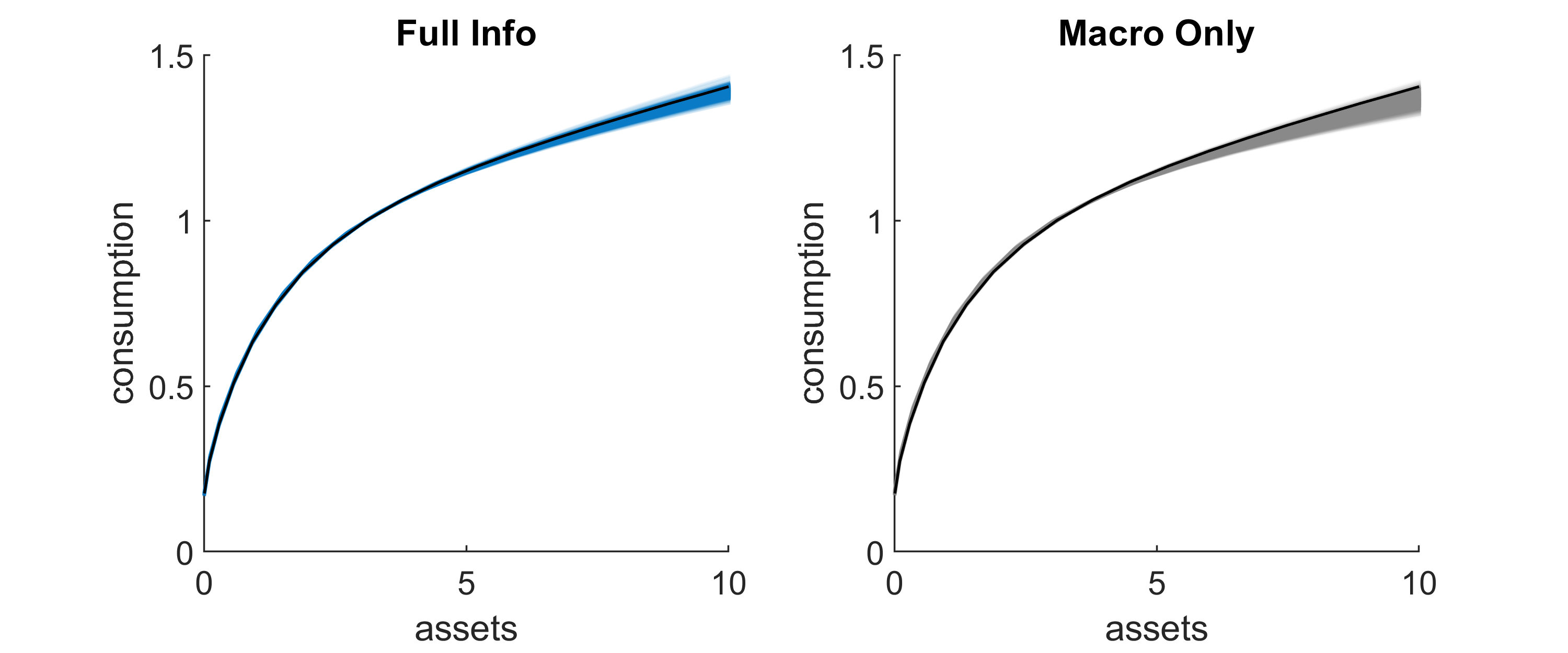} 
\caption{Posterior draws of steady-state consumption policy function for unemployed households. See caption for \cref{fig:example_hh_consumption}.}
\label{fig:example_hh_consumption_unemp}
\end{figure}

\begin{figure}[tp]
\centering
\textsc{Het.\ household model: Impulse responses of asset distribution, unemployed} \\[\baselineskip]
\includegraphics[width=\linewidth,clip=true,trim=3em 0 3em 0]{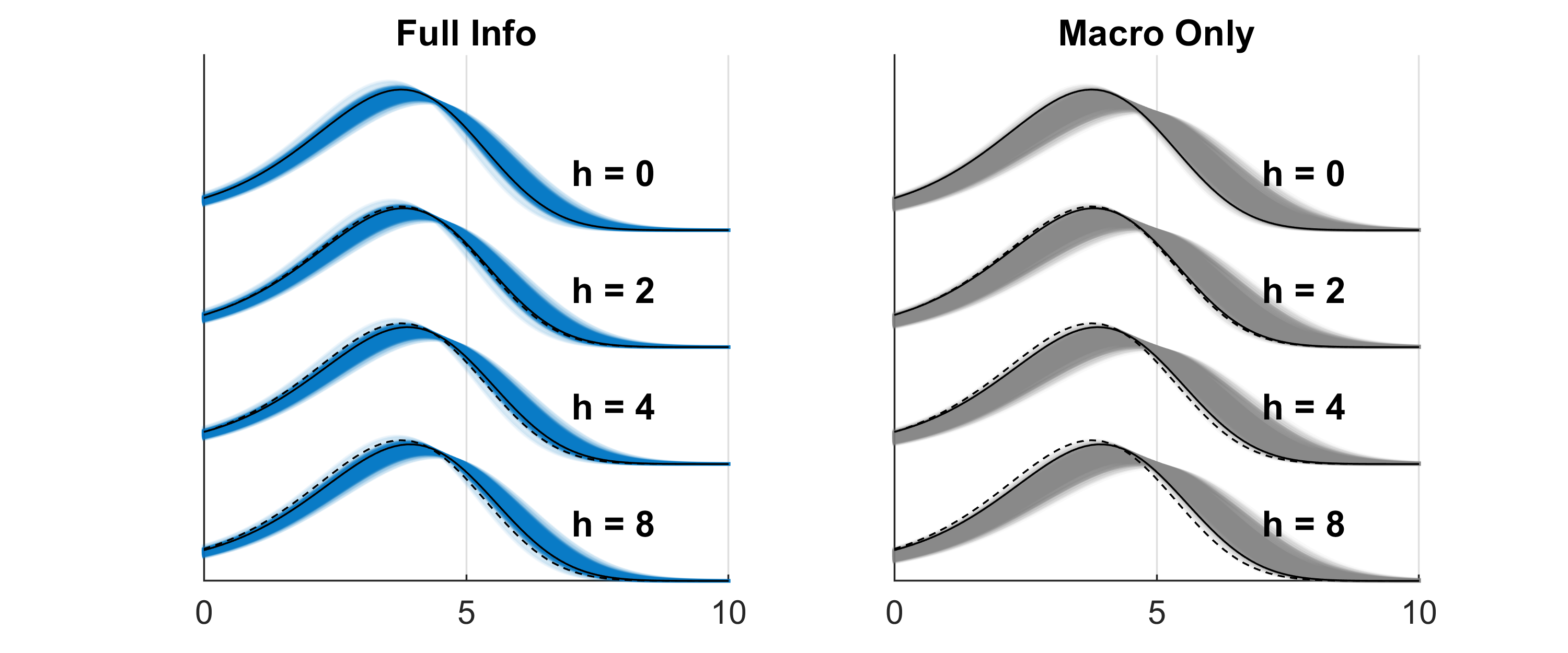} 
\caption{Posterior of impulse response function of unemployed households' asset distribution with respect to an aggregate productivity shock. See caption for \cref{fig:example_hh_irf}.}
\label{fig:example_hh_irf_unemp}
\end{figure}

Here we provide additional results for the numerical illustration of the heterogeneous household model. \cref{fig:example_hh_consumption_unemp} shows the full-information and macro-only posterior distributions of the steady-state consumption policy function for unemployed households. \cref{fig:example_hh_irf_unemp} shows the full-information and macro-only posterior distributions of the impulse response function of the asset distribution for unemployed households with respect to a TFP shock. In terms of the comparison between full-information and macro-only inference, both these figures are qualitatively similar to those for employed households, cf.\ \cref{fig:example_hh_consumption,fig:example_hh_irf} in the main paper.

\begin{figure}[tp]
\centering
\textsc{Heterogeneous household model: Posterior density, $N=100$} \\[\baselineskip]
\includegraphics[width=\linewidth]{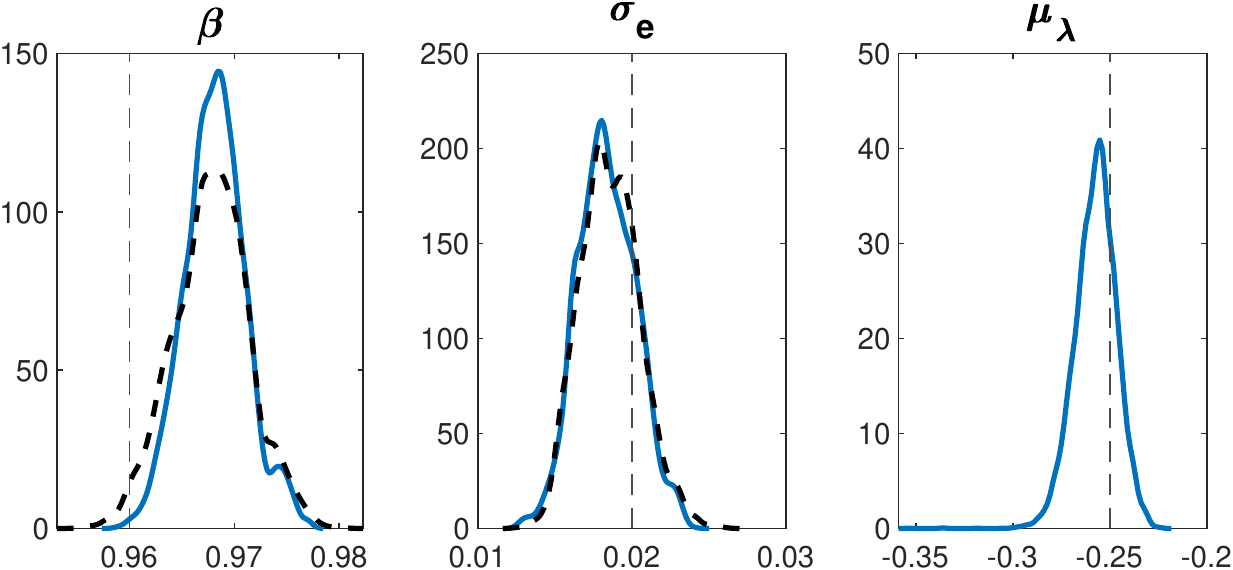}
\caption{Posterior densities with (blue solid curves) and without (black dashed curves) conditioning on the micro data, for cross-sectional sample size $N=100$. See caption for \cref{fig:example_hh_postdens}.}
\label{fig:example_hh_postdens_N100}
\end{figure}

\cref{fig:example_hh_postdens_N100} shows the full-information and macro-only posterior densities of the model parameters in an alternative simulation where we only observe $N=100$ micro draws every ten periods (instead of $N=1000$). All other settings are the same as in \cref{sec:example_hh}. Naturally, the accuracy of posterior inference is affected by the smaller sample size, but we see that the individual heterogeneity parameter $\mu_\lambda$ is still precisely estimated in this simulation.

\clearpage

\section{Heterogeneous firm model}
\subsection{Model assumptions}
\label{sec:example_firm_details}
We here briefly describe the assumptions of the heterogeneous firm model. See \citet{Khan2008} and \citet{Winberry2018} for more complete discussions of the model. Note that the notation in this section recycles some of the notation used for the household model in \cref{sec:model_hh}.

A unit mass of firms $i \in [0,1]$ have decreasing returns to scale production functions $Y_{i,t} = e^{\zeta_t+\epsilon_{i,t}}k_{i,t}^\alpha n_{i,t}^\nu$, where $k_{i,t}$ and $n_{i,t}$ denote firm-specific capital and labor inputs ($\alpha+\nu<1$). Labor $n_{i,t}$ is hired in a competitive labor market with aggregate wage rate $w_t$. Aggregate log TFP $\zeta_t$ evolves as an AR(1) process $\zeta_t = \rho_\zeta \zeta_{t-1} + \varepsilon_{\zeta,t}$, $\varepsilon_{\zeta,t} \overset{i.i.d.}{\sim} N(0,\sigma_\zeta^2)$. The firm-specific log productivity levels evolve as independent AR(1) processes $\epsilon_{i,t} = \rho_\epsilon \epsilon_{i,t-1} + \omega_{i,t}$, where the idiosyncratic shocks $\omega_{i,t} \overset{i.i.d.}{\sim} N(0,\sigma_\epsilon^2)$ are independent across $i$ and are dynamically independent of aggregate TFP.

After production, firms can choose to turn part of their production good into investment in next-period capital. A gross investment level of $I_{i,t}$ yields next-period capital $k_{i,t+1} = (1-\delta)k_{i,t} + e^{q_t} I_{i,t}$. The aggregate investment efficiency shifter $q_t$ follows an AR(1) process $       q_t = \rho_q q_{t-1} + \varepsilon_{q,t}$, where the aggregate shock $\varepsilon_{q,t} \overset{i.i.d.}{\sim} N(0,\sigma_q^2)$ is independent of the aggregate TFP shock $\varepsilon_{\zeta,t}$. Investment activity is free if $|I_{i,t}/k_{i,t}| \leq a$, where $a \geq 0$ is a parameter. Otherwise, firms pay a fixed adjustment cost of $\xi_{i,t}$ in units of labor (i.e., the monetary cost is $\xi_{i,t} \times w_t$). $\xi_{i,t}$ is drawn at the beginning of every period from a uniform distribution on the interval $[0, \bar{\xi}]$, independently across firms and time. Here $\bar{\xi} \geq 0$ is another parameter.

A representative household chooses consumption $C_t$ and labor supply $L_t$ to maximize
\[\mathbb{E}_0\left[\sum_{t=0}^\infty \beta^t \left\{ \log C_t - \chi \frac {L_t^{1+\varphi}} {1+\varphi} \right\} \right],\]
where $\varphi$ is the inverse Frisch elasticity of labor supply. The household owns all firms and markets are complete. Market clearing requires $C_t = \int (Y_{i,t} + I_{i,t})\,di$ and $L_t = \int n_{i,t}\, di$.

See \citet[section 2.2]{Winberry2018} for the Bellmann equations implied by the firms' and household's optimality conditions.

\subsection{Calibration}
\label{sec:example_firm_calib}
\cref{tab:example_firm_calib} shows the parameter calibration used to simulate the data. The labor disutility parameter $\chi$ is chosen so that steady-state hours equal $\bar{N}=1/3$, given all other parameters. As explained in \cref{sec:example_firm_baseline}, the only difference from \citet{Winberry2018} is that the idiosyncratic productivity process uses the alternative (less persistent) parametrization from \citet{Khan2008}. We do this because \citeauthor{Winberry2018}'s Dynare code appears to be more numerically stable in a neighborhood of these alternative parameter values.

\begin{table}[tp]
\centering
\textsc{Heterogeneous firm model: Parameter calibration}  \\[\baselineskip]
\renewcommand{\arraystretch}{1.1}
\begin{tabular}{lll|lll}
\hline
$\beta$         &         Discount factor       &       0.961   &       $\overline\xi$         &        Fixed cost bound       &       0.0083  \\
$\chi$  &        Labor disutility       &        $\bar{N}=\frac 1 3$    &        $\rho_\zeta$    &        Agg.\ TFP AR(1)        &       0.859   \\
$\varphi$       &       Inverse Frisch  &       $10^{-5}$ &         $\sigma_\zeta$         &        Agg.\ TFP AR(1)        &       0.014   \\
$\nu$   &        Labor share    &       0.64    &         $\rho_q$      &        Agg.\ inv.\ eff.\ AR(1)         &       0.859   \\
$\alpha$        &        Capital share  &       0.256   &         $\sigma_q$         &        Agg.\ inv.\ eff.\ AR(1)        &       0.014   \\
$\delta$        &        Capital depreciation   &       0.085   &         $\rho_\epsilon$        &        Idio.\ TFP AR(1)       &       0.53    \\
$a$     &        No fixed cost region   &       0.011   &         $\sigma_\epsilon$         &        Idio.\ TFP AR(1)       &       0.0364  \\

          \hline             
                \end{tabular}
\caption{Parameter calibration in the heterogeneous firm model.} \label{tab:example_firm_calib}
\end{table}

\subsection{Estimating the parameters of the firms' productivity process}
\label{sec:example_firm_prod}
In this subsection we run the same estimation exercise as in \cref{sec:example_firm}, except that we here estimate the AR(1) parameter $\rho_\epsilon$ and innovation standard deviation $\sigma_\epsilon$ of the firms' idiosyncratic log productivity process. All other structural parameters (including the adjustment cost parameters) are assumed known for simplicity. The data and estimation settings are similar to those in \cref{sec:example_firm_results} except that micro cross sections are observed at each of the five time points $t=10,20,\dots,50$.

\begin{figure}[tp]
\centering
\textsc{Heterogeneous firm model: Posterior densities of productivity parameters} \\[\baselineskip]
\includegraphics[width=\linewidth]{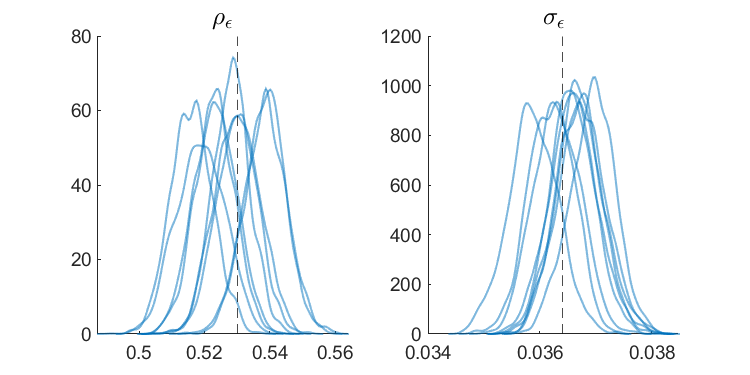}
\caption{Posterior densities across 9 simulated data sets. Vertical dashed lines indicate true parameter values. Posterior density estimates from the 9,000 retained MCMC draws using Matlab's {\tt ksdensity} function with default settings.}
\label{fig:example_firm_prod_postdens}
\end{figure}

\cref{fig:example_firm_prod_postdens} shows the full-information posterior densities of the idiosyncratic productivity parameters $(\rho_\epsilon,\sigma_\epsilon)$, across 9 simulated data sets.\footnote{We simulated 10 data sets but discarded one, as standard MCMC convergence diagnostics showed a failure of convergence of the Metropolis-Hastings sampler for that particular data set.} The posterior densities are well-centered and concentrated near the true parameter values. We refrain from comparing with posterior inference that only exploits macro data, as these posteriors are extremely diffuse, consistent with \citet{Khan2008}.

\section{Alternative panel data approach} \label{sec:panel_general}

We here propose an alternative approach to handling panel data that may sometimes be more convenient than the procedure discussed in \cref{sec:panel}. Another view of the challenge described in that section is that, while the solution to the heterogeneous agent model supplies the marginal cross-sectional distribution of micro state variables at any point in time $p\left(s_{i,t}\mid z_{t},\theta\right)$, it does not directly supply the \emph{joint} density of micro state variables across multiple points in time $p\left(\left\{s_{i,t}\right\}_{t\in\mathcal{T}}\mid \mathbf{z},\theta\right)$. To get around this issue, our second proposal is to artificially expand the micro state vector when solving the heterogeneous agent model, by including the previous periods' state variables in the current state vector.

Again, consider for illustration the heterogeneous household model in \cref{sec:model_hh}, and suppose each household is observed for two consecutive periods. Rather than treating the micro state vector as simply $s_{i,t}=(\lambda_i,\epsilon_{i,t},a_{i,t-1})$ (permanent productivity, current-period employment, and predetermined normalized assets), we now expand it to be $\tilde s_{i,t}=(\lambda_i,\epsilon_{i,t},a_{i,t-1},\epsilon_{i,t-1},a_{i,t-2})$ (thus including previous-period employment and normalized assets). Let $\tilde\psi_t$ denote the parameters governing the full four-dimensional cross-sectional distribution for the expanded micro state variables $\tilde s_{i,t}$. Then $\tilde\psi_t$ is part of the expanded aggregate state  $\tilde z_t$.

Having expanded the micro state vector, we now modify the numerical model solver so that it outputs the joint cross-sectional density of $\tilde{s}_{i,t}$, not just of $s_{i,t}$. This is achieved by altering the model solution code and continuing to impose all model-implied restrictions on the evolution of the cross-sectional distribution (which is summarized by distribution parameters $\tilde\psi_t$). Note that the model-implied restrictions include those implied by the individual saving behavior $a_{t}'(a,\epsilon)$ and the exogenous Markov process for employment $p(\epsilon_{i,t} \mid \epsilon_{i,t-1},\theta)$. We also continue to apply the \citet{Reiter2009} type model solution method, leading to a state space model for macro variables characterized by equations \eqref{eqn:loglin_transition} and
\eqref{eqn:loglin_meas}.

Given a draw of the macro states $\mathbf{\tilde z}$ outputted from the modified model solution, we can easily compute the micro likelihood for two consecutive periods of household employment and income $(y_{i,t},y_{i,t-1})=(\epsilon_{i,t},\iota_{i,t},\epsilon_{i,t-1},\iota_{i,t-1})$. This can be calculated as a simple distributional transformation of the four-dimensional distribution of the expanded micro state vector $\tilde s_{i,t}$, using the definition $\iota_{i,t'} = \lambda_i\lbrace w_{t'}[(1-\tau)\epsilon_{i,t'} + b(1-\epsilon_{i,t'})] + (1+r_{t'})a_{i,t'-1} \rbrace$ for $t' \in \lbrace t-1,t\rbrace$.

Though conceptually simple, the downside of the approach described in this section is that the expansion of the micro state vector is computationally demanding, for two reasons. First, as the dimension of the effective micro state vector $\tilde s_{i,t}$ increases, so does the dimension of the distribution parameters $\tilde \psi_{t}$, and then both the speed and the precision of off-the-shelf numerical methods for solving heterogeneous agent models tend to deteriorate. Second, achieving a sufficiently accurate approximation of the cross-sectional distribution of the higher-dimensional expanded micro state vector $\tilde s_{i,t}$ may require using a large number $q$ of basis functions in the finite-dimensional distributional approximation described in \cref{sec:model_hh}, which further increases the dimension of $\tilde \psi_t$ in the equilibrium approximation.

\end{appendices}

\clearpage

\phantomsection
\addcontentsline{toc}{section}{References}
\bibliography{het_agents_ref}